\documentclass[aip,pop,reprint,superscriptaddress,longbibliography]{revtex4-2}

\usepackage[T1]{fontenc}
\usepackage[utf8]{inputenc}
\usepackage{lmodern}

\usepackage{amsmath}
\usepackage{amssymb}
\usepackage{amsfonts}
\usepackage{amsthm}
\usepackage{mathrsfs}
\usepackage{bm}

\usepackage{graphicx}
\usepackage[dvipsnames]{xcolor}
\usepackage{booktabs}
\usepackage{array}

\usepackage{amsfonts}
\usepackage{babel}
\usepackage{enumitem}
\usepackage[normalem]{ulem}
\usepackage{soul}
\usepackage{subcaption}
\usepackage{makecell}
\usepackage[percent]{overpic}
\usepackage{comment}

\usepackage{float}

\usepackage{caption}
\usepackage{placeins}

\usepackage[dvipsnames]{xcolor}
\usepackage{tikz}
\usepackage{hyperref}
\hypersetup{
    breaklinks = true,
    colorlinks = true,
    citecolor = {blue},
    urlcolor = {blue},
    linkcolor = {blue}
}

\usetikzlibrary{positioning,arrows.meta,shapes.geometric,calc,fit,backgrounds}

\tikzstyle{io} = [trapezium, 
trapezium stretches=true, 
trapezium left angle=80, 
trapezium right angle=100, 
minimum width=3cm, 
minimum height=1cm, 
text centered, 
text width=4.1cm, 
draw=black,
line width=0.8,
]

\tikzstyle{process} = [rectangle, 
minimum width=4cm, 
minimum height=1cm, 
text centered, 
text width=82cm, 
draw=black,
line width=0.8,
]

\tikzstyle{decision} = [diamond, 
minimum width=3cm, 
minimum height=1cm, 
text centered, 
aspect=2,
draw=black,
line width=0.8,
]

\tikzstyle{startstop} = [rectangle, 
rounded corners, 
minimum width=3cm, 
minimum height=1cm,
text centered, 
draw=black,
line width=0.8,
]

\tikzstyle{arrow} = [ultra thick,->,>=stealth]

\tikzstyle{dash_arrow} = [ultra thick,->,>=stealth, dashed]

\begin{document}

\title{ELMO: An Uncertainty-Aware Simulation-to-Surrogate Workflow for Fast Pedestal Linear-Stability Prediction}


\author{Nami Li}
\email{li55@llnl.gov}
\affiliation{
Lawrence Livermore National Laboratory,
Livermore, California 94550, USA
}

\author{X. Q. Xu}
\affiliation{
Lawrence Livermore National Laboratory,
Livermore, California 94550, USA
}

\author{T. Osborne}
\affiliation{
General Atomics,
San Diego, California 92186, USA
}

\author{E. Suchyta}
\affiliation{
Oak Ridge National Laboratory,
Oak Ridge, Tennessee 37831, USA
}

\author{Y. C. Fu}
\affiliation{
Columbia University,
New York, New York 10027, USA
}

\author{N. Podhorszki}
\affiliation{
Oak Ridge National Laboratory,
Oak Ridge, Tennessee 37831, USA
}

\author{H. Wang}
\affiliation{
General Atomics,
San Diego, California 92186, USA
}

\author{Z. Li}
\affiliation{
General Atomics,
San Diego, California 92186, USA
}


\date{\today}


\begin{abstract}
Rapid prediction of pedestal linear stability is important for efficient exploration of tokamak operating space, uncertainty quantification, and future model-informed plasma control. However, mode-resolved magnetohydrodynamic stability calculations using \textsc{BOUT++} remain computationally expensive. We present a focused implementation of ELMO---the Edge Learning and Modeling Orchestrator---as an uncertainty-aware simulation-to-surrogate workflow integrating structured equilibrium generation, field-aligned mesh construction, large-scale \textsc{BOUT++} linear-stability calculations, automated campaign execution and data reduction, and Gaussian Process Regression (GPR).

For a single DIII-D plasma shape, 3,869 of 7,992 requested equilibrium configurations completed the equilibrium-reconstruction, mesh-generation, stability-calculation, and quality-control stages. Each retained equilibrium was evaluated at sixteen toroidal mode numbers, $n=5$--80 with $\Delta n=5$, using ideal-MHD and ideal-plus-diamagnetic physics models. The resulting database contains 123,808 mode-resolved \textsc{BOUT++} calculations. Using eight equilibrium-derived pedestal features, the GPR surrogate predicts two sixteen-mode growth-rate spectra together with latent posterior uncertainty estimates.

Independent-test performance was evaluated separately for five random data-partition realizations and summarized by the mean and sample standard deviation. For the maximum linear growth rate, the ideal-MHD model achieved $R^2=0.978\pm0.013$, while the ideal-plus-diamagnetic model achieved $R^2=0.966\pm0.009$. The surrogate also accurately reproduces the overall growth-rate spectral shape and dominant unstable toroidal mode. Independent-test calibration diagnostics show that the latent posterior uncertainties provide useful relative acquisition scores but are underdispersed, particularly for the ideal-plus-diamagnetic model, and should not be interpreted as fully calibrated prediction intervals.
A complete prediction of all 32 output quantities requires approximately 20~ms on a single CPU core, compared with approximately 21~min using 128 CPU cores for the corresponding \textsc{BOUT++} scan, yielding an approximately $6.3\times10^{4}$-fold wall-clock speedup and an approximately $8.1\times10^{6}$-fold reduction in computational cost. Pool-based uncertainty-guided selection provides a modest improvement in sample efficiency over intermediate training-set sizes, although its performance converges with random selection as the available pool is exhausted. This implementation establishes a rapid simulation-to-surrogate capability and a foundation for future multi-shape, nonlinear, and closed-loop adaptive workflows.
\end{abstract}

\maketitle


\section{Introduction}

The edge transport barrier formed during high-confinement-mode (H-mode) operation plays a central role in determining the performance of magnetically confined fusion plasmas. The steep pressure gradients and associated bootstrap current within the pedestal can destabilize peeling--ballooning and related edge magnetohydrodynamic (MHD) instabilities, which may develop into edge-localized modes (ELMs). Large ELMs periodically expel particles and energy from the confined plasma and impose transient heat loads on plasma-facing components. Reliable assessment of pedestal stability is therefore important for scenario development, ELM mitigation, and the operation of future burning-plasma devices, including ITER and SPARC~\cite{shimada2007overview,casper2013development,creely2020overview,hughes2020projections,wagner2007quarter,zohm1996edge,leonard2014edge,pitts2019physics}.

Substantial progress has been made in physics-based modeling of pedestal structure and stability. The EPED model combines pedestal transport constraints with peeling--ballooning stability theory to predict pedestal height and width self-consistently~\cite{snyder2011eped}. Linear stability is commonly evaluated using ideal-MHD eigenvalue solvers such as ELITE~\cite{wilson2002elite}, while extended-MHD codes such as \textsc{BOUT++} provide a flexible framework for linear and nonlinear studies of edge instabilities with additional physical effects~\cite{dudson2009bout,xu2010nonlinear,li2014linear}. These approaches have substantially advanced the understanding of pedestal stability, but systematic parameter studies remain computationally demanding because they require large ensembles of self-consistent equilibria, field-aligned computational meshes, and mode-resolved stability calculations.

Machine-learning surrogates offer a complementary approach for accelerating repeated stability evaluations. Neural-network models have reproduced pedestal predictions from theory-based frameworks such as EPED and EuroPED~\cite{meneghini2017self,alvarez2024europed}. Surrogates trained on MISHKA-generated databases have also demonstrated rapid prediction of peeling--ballooning stability boundaries for KSTAR and JET~\cite{mikhailovskii1997optimization,heo2023neural,heo2024development,bruncrona2025machine}. These studies establish the potential of scientific machine learning for pedestal analysis. However, surrogate reliability depends on the coverage, consistency, and quality of the training database. This dependence creates a particular challenge when each training sample requires an expensive physics simulation and when surrogate uncertainty must be assessed in sparsely sampled regions~\cite{hestness2017deep,bahri2024explaining}.

These considerations motivate integrated simulation-to-surrogate workflows that connect systematic equilibrium generation, automated simulation execution, reduced-data extraction, quality control, and uncertainty-aware learning. Such workflows should preserve the associations among equilibrium parameters, computational meshes, physics configurations, and stability outputs while producing compact data products suitable for surrogate development. Probabilistic surrogate models are especially attractive in this setting because their model-based posterior uncertainties can help identify comparatively under-represented regions and guide the selection of additional high-fidelity calculations~\cite{rasmussen2006gaussian}.

Within the broader ABOUND research program, \textbf{ELMO---the Edge Learning and Modeling Orchestrator}---is being developed as an integrated simulation-to-surrogate framework for boundary-plasma modeling. The present work reports a focused ELMO implementation for rapid prediction of linear pedestal stability. The workflow integrates systematic equilibrium generation using \textsc{Varyped}~\cite{Osborne_2015}/BOUT\_DB, field-aligned mesh generation using \textsc{Hypnotoad}~\cite{hypnotoad_docs}, mode-resolved \textsc{BOUT++} linear-stability calculations~\cite{dudson2009bout}, automated campaign execution and reduced-data extraction, physics-informed feature selection, and Gaussian Process Regression (GPR)~\cite{michoski2024gaussian}.

The principal contribution of this work is the integration and quantitative evaluation of these components as a provenance-preserving, campaign-scale simulation-to-surrogate workflow. The implementation maintains traceable associations among equilibrium-generation parameters, computational meshes, BOUT++ physics configurations, execution status, reduced stability outputs, and surrogate-training data. Its predictive performance is assessed using equilibrium-level separation of adaptive-training, validation, and independent-test subsets, with the complete procedure repeated over five random data-partition realizations. Gaussian Process posterior uncertainties are evaluated through independent-test calibration diagnostics and are used as relative acquisition scores rather than assumed to constitute calibrated prediction intervals. Together with explicit measurements of prediction time and computational cost, these elements establish the accuracy, uncertainty interpretation, and practical computational benefit of the integrated workflow.

The quality-controlled database contains 3,869 equilibria. For each equilibrium, linear growth rates are calculated for sixteen toroidal mode numbers using both ideal-MHD and ideal-plus-diamagnetic physics models, yielding two sixteen-mode growth-rate spectra and 32 surrogate output quantities. The resulting database therefore contains 123,808 mode-resolved \textsc{BOUT++} linear-stability calculations. Using eight equilibrium-derived pedestal features, the Gaussian Process surrogate reproduces the growth-rate spectra together with the maximum linear growth rate and dominant unstable toroidal mode. A complete surrogate prediction requires approximately 20~ms on a single CPU core, compared with approximately 21~min for the corresponding \textsc{BOUT++} scan using 128 CPU cores, corresponding to an approximately $6.3\times10^{4}$-fold wall-clock speedup.

The adaptive component demonstrated in this work is restricted to pool-based sample selection within the existing simulation database. Gaussian Process posterior uncertainty is used to rank candidate equilibria for inclusion in the training set, but no new \textsc{Varyped} equilibria, \textsc{Hypnotoad} meshes, or \textsc{BOUT++} calculations are generated automatically. The quality-controlled database is partitioned at the equilibrium level into an adaptive-training pool (70\%), a validation set (15\%), and an independent test set (15\%). The validation set is used for feature selection, learning-curve monitoring, and adaptive stopping, whereas the independent test set remains excluded from all model-development decisions and is used only for final performance evaluation. Consequently, the present work demonstrates an uncertainty-aware simulation-to-surrogate capability for single-shape linear pedestal-stability prediction rather than a completed closed-loop adaptive simulation framework. Extensions to multiple plasma shapes, additional linear and nonlinear physics models, automated generation of new simulations, and experimental validation remain subjects for future work.

The remainder of this paper is organized as follows. Section~\ref{sec:workflow} describes the ELMO simulation-to-surrogate workflow and construction of the quality-controlled \textsc{BOUT++} database. Section~\ref{sec:workflow_ml} presents the Gaussian Process surrogate methodology, including the regression formulation, physics-informed feature selection, separate validation and independent-test subsets, and uncertainty-guided pool-based adaptive sample selection. Section~\ref{sec:statistics} characterizes the equilibrium-parameter coverage and linear-stability properties of the simulation database. Section~\ref{sec:verification} evaluates feature-selection performance, surrogate convergence, adaptive sample-selection performance, prediction accuracy, uncertainty calibration, and computational efficiency. Section~\ref{sec:discussion} discusses the interpretation, domain of validity, and principal limitations of the present implementation. Section~\ref{sec:conclusion} summarizes the main conclusions.

\section{ELMO Simulation-to-Surrogate Workflow}
\label{sec:workflow}

\begin{figure*}[t]
    \centering
    \begin{tikzpicture}[node distance=1.6cm, font=\footnotesize]
    
        \node (in1) [io, text width=3.65cm] {
        \textbf{Initial pedestal parameters sample} \\
        $p_\mathrm{ped}$, $\Delta_\mathrm{ped}$, $R_\mathrm{p,peak}$, $J_\mathrm{b}$, ...
        };
        
        \node (pro1) 
        [process, below of=in1, text width=3.65cm, yshift=-1.8cm]
        {
        \textbf{Varyped database:} 2-D equilibria\\
        Extend 1-D Profile from pedestal to SOL with EFIT       
        \vspace{0.1cm}
        \includegraphics[width=3.6cm]{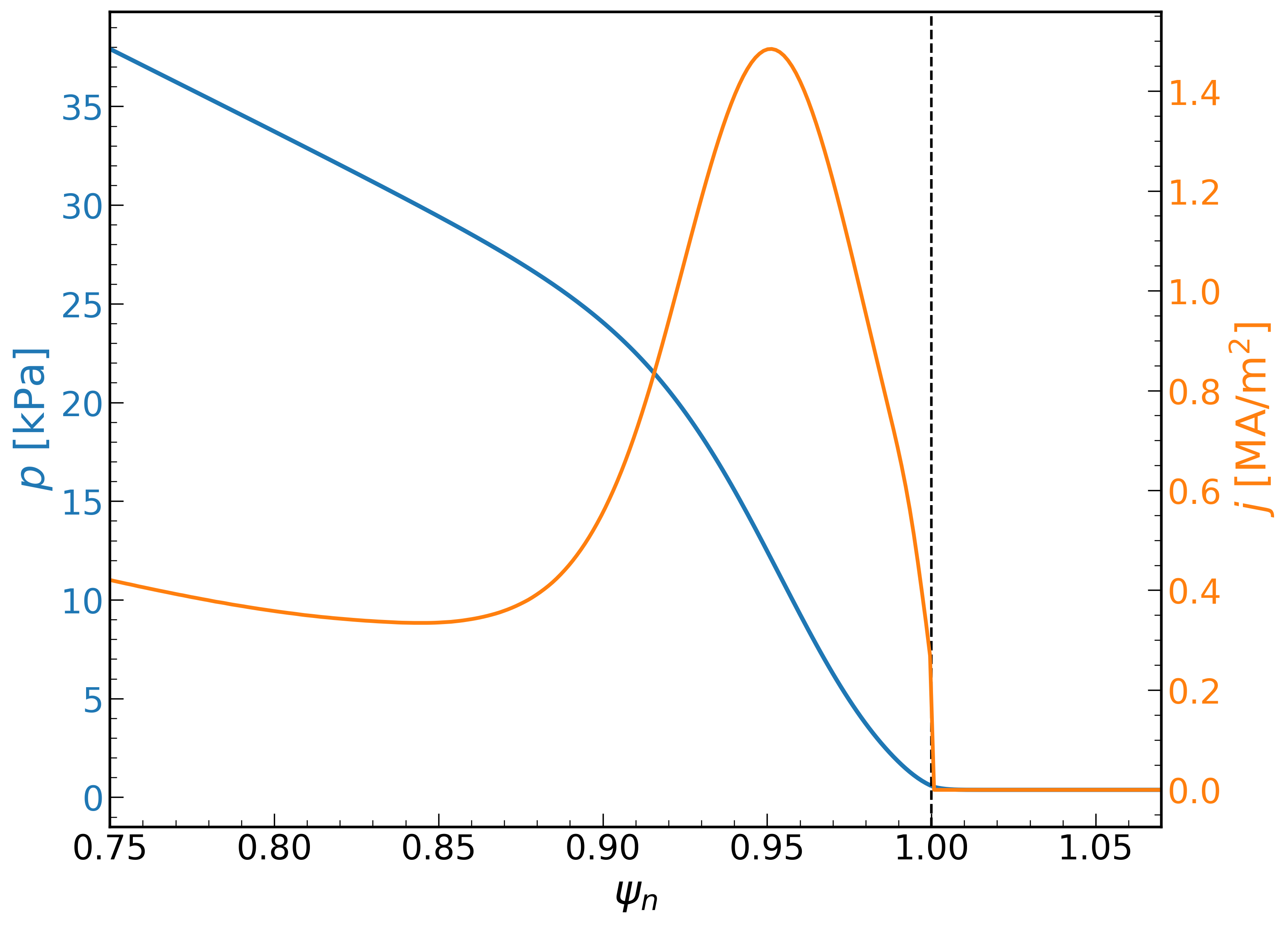}
        };
        \draw [arrow] (in1) -- (pro1);

        \node (pro2) 
        [process, right of=pro1, text width=3.4cm, minimum height=6.25cm, xshift=2.9cm, yshift=1cm] 
        {
        \textbf{Hypnotoad}: BOUT++ grid generation 
        
        \vspace{0.1cm}
        \includegraphics[width=3.2cm]{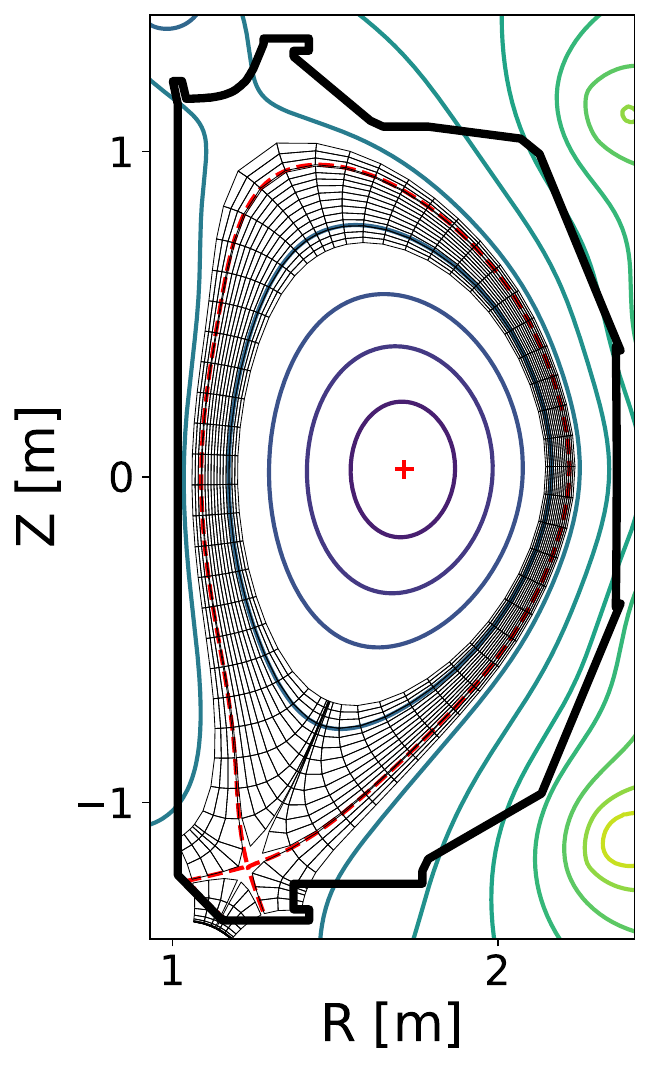}
        };
        \draw [arrow] (2.,-2.4) -- (2.5,-2.4);

        \node (pro3) 
        [process, right of=pro2, text width=7.5cm, minimum height=6.25cm,
        xshift=4.75cm]
        {
        \textbf{Automated BOUT++ Campaign \\ and reduced database}
        \vspace{0.1cm}
        
        \includegraphics[width=7cm]{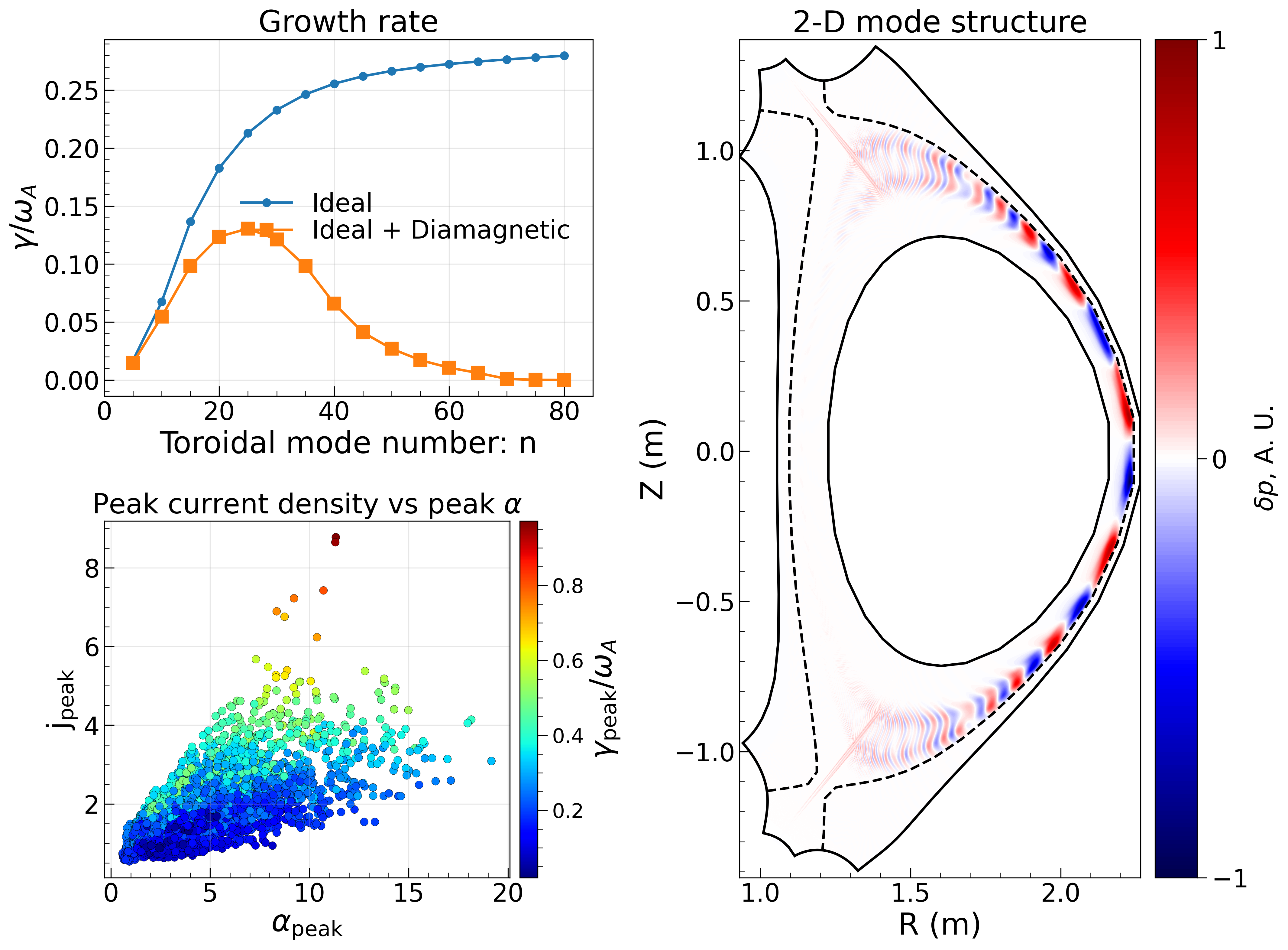}
        };
        \draw [arrow] (6.5,-2.5) -- (6.99,-2.5);

        \begin{scope}[on background layer]
        \node[draw=green!70!black, dashed, line width=1.5pt, rounded corners=4pt, inner sep=8pt, fit=(in1)(pro1)(pro2)(pro3),
        label={[text=green!70!black,font=\bfseries]north:Simulation Data Campaign}] {};
        \end{scope}

        \node (pro4) 
        [process, below of=pro3, 
        minimum width=3cm, 
        minimum height=1.5cm,
        text width=4cm,
        xshift=1.7cm,
        yshift=-4cm] 
        {
        \textbf{Train AI/ML surrogate models}
        \vspace{0.1cm}

        \includegraphics[width=3.5cm]{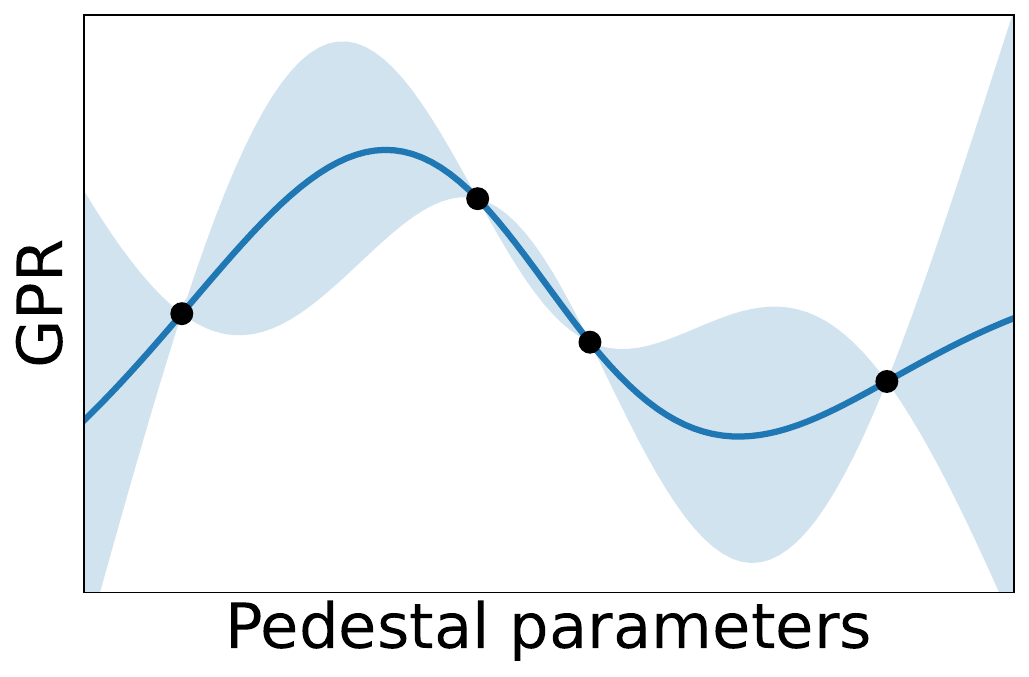}
        };
        \draw [arrow] (12.55, -5.7) -- (pro4);
        
        \node (dec1) 
        [decision, left of=pro4, 
        minimum width=2cm, 
        minimum height=1.5cm,
        text width=2.3cm,
        xshift=-4.75cm
        ] 
        {\textbf{Ready to exit?}};
        \draw [arrow] (pro4) -- (dec1);

        \node (stop1) 
        [startstop, below of=dec1, text width=2.45cm, 
        yshift=-0.5cm] 
        {
        \textbf{Return surrogates}
        };
        \draw [arrow] (dec1) -- node[anchor=west] {Yes} (stop1);

        \node (pro5) 
        [process, 
        below of=pro1,text width=4.cm, 
        minimum width=4.cm, 
        minimum height=1.5cm,
        text width=4cm,
        yshift=-3.cm] 
        {
        \textbf{Sample selection:} \\ Add new samples using active learning strategies.};
        
        \draw [arrow] (dec1) -- node[anchor=south] {No} (pro5);
       \draw [dash_arrow, orange!90!black] (pro5.north) -- node[ anchor=west, text width=2.8cm, align=left, xshift=1mm, yshift=-2mm, text=black]
       {\textcolor{orange!90!black}{Future closed-loop\\database expansion}\\
       {\footnotesize (new simulations)}} 
       (0., -5.55);
        \draw[arrow] (pro5.south) -- ++(0,-2cm) -| (pro4.south);
        \node[font=\footnotesize,anchor=north, text=green!50!black]
        at ($(pro5.south)!0.5!(pro4.south)+(0,-1.6cm)$)
        {Update the training subset and retrain AI/ML model using existing simulations};

    \end{tikzpicture}
    \caption{\textbf{Overview of the ELMO simulation-to-surrogate workflow.}} 
    \label{fig:ELMO_workflow}
\end{figure*}

\subsection{Scope and Overall Workflow}

\textbf{ELMO---the Edge Learning and Modeling Orchestrator---} is being developed within the broader ABOUND research program to connect boundary-plasma simulations, automated workflow execution, reduced-data generation, surrogate modeling, uncertainty quantification, and adaptive case selection. The present study demonstrates a focused implementation for rapid prediction of linear pedestal stability using a structured DIII-D-based equilibrium database, mode-resolved \textsc{BOUT++} stability calculations, and Gaussian Process Regression (GPR).

Figure~\ref{fig:ELMO_workflow} summarizes the simulation-to-surrogate pathway. A structured pedestal scan is first generated using \textsc{Varyped}/BOUT\_DB while retaining a single DIII-D plasma boundary shape. Converged equilibria are converted into field-aligned computational meshes using \textsc{Hypnotoad} and evaluated with \textsc{BOUT++} using two linear physics models over a prescribed set of toroidal mode numbers. Automated campaign tools manage the calculations and reduce their outputs to quality-controlled growth-rate spectra, stability metrics, and equilibrium-derived features. These data are subsequently used to train, validate\textcolor{red}{,} and independently test an uncertainty-aware GPR surrogate.

The adaptive component demonstrated here is restricted to pool-based sample selection from the existing simulation database. Gaussian Process posterior uncertainty is used to rank candidate equilibria for inclusion in the training set; no new \textsc{Varyped} equilibria, \textsc{Hypnotoad} meshes, or \textsc{BOUT++} calculations are generated during this procedure. The demonstrated scope is therefore limited to a single-shape, linear-stability simulation-to-surrogate workflow. Closed-loop adaptive simulation, multiple plasma shapes, additional linear and nonlinear physics models, and experimental validation remain subjects for future work.

\subsection{Varyped Equilibrium Generation}

The equilibrium and profile database used in this study was generated using the DIII-D \textsc{Varyped} workflow. \textsc{Varyped}~\cite{Osborne_2015} was originally developed to construct controlled pedestal-equilibrium databases for peeling--ballooning stability calculations with ELITE~\cite{wilson2002elite} and for comparisons with EPED~\cite{snyder2011eped}-type pedestal models. Within the present ELMO implementation, \textsc{Varyped}/BOUT\_DB provides the equilibrium- and profile-generation layer connecting a structured pedestal-parameter scan to Hypnotoad mesh generation, BOUT++ stability calculations, and surrogate-model development.

Starting from an experimentally reconstructed DIII-D kinetic equilibrium, \textsc{Varyped} systematically modifies parameterized pressure and current-density profiles and reconstructs a self-consistent EFIT~\cite{Lao_1985} equilibrium for each requested parameter combination. The pressure- and current-profile controls can be varied independently, allowing systematic exploration of peeling--ballooning stability space while maintaining the force-balance and magnetic-geometry information required by BOUT++. An iterative process is used to construct an equilibrium consistent with the specified pressure and current density profiles from the initial experimental equilibrium. 

The present database was generated using a structured multidimensional scan over prescribed discrete parameter values rather than uncertainty-guided sampling for this study. The scan varies the pedestal-pressure controls, pedestal-current controls, and edge safety factor while retaining a single prescribed DIII-D plasma boundary shape. The pressure and current functional forms and their corresponding scan controls are described below.

\subsubsection{Pressure-profile parameterization}

The pressure profile is represented using the modified hyperbolic-tangent form employed by \textsc{Varyped},

\begin{equation}
p(\psi_N)
=
\frac{p_{\mathrm{ped}}-p_{\mathrm{foot}}}{2}
\left[
\frac{(1+cz)e^z-e^{-z}}
     {e^z+e^{-z}}
\right]
+
\frac{p_{\mathrm{ped}}+p_{\mathrm{foot}}}{2},
\label{eq:varyped-pressure}
\end{equation}

where

\begin{equation}
z
=
\frac{2(\psi_{p,\mathrm{sym}}-\psi_N)}
     {p_{\mathrm{wid}}}.
\label{eq:varyped-pressure-z}
\end{equation}

Here, $\psi_N$ is the normalized poloidal flux,
$p_{\mathrm{ped}}$ and $p_{\mathrm{foot}}$ are the pedestal top and foot pressures, $p_{\mathrm{wid}}$ is the pedestal width, and $\psi_{p,\mathrm{sym}}$ denotes the location of the maximum pressure gradient. The parameter $c$ controls the slope of the core-pressure profile and is adjusted to satisfy the requested overall normalized beta. In the EFIT~\cite{Lao_1985} equilibrium solver only the pressure gradient as a function of $\psi_N$ is specified. Although the pressure gradient can be non-zero at the separatrix, the fixed boundary version of EFIT requires the pressure at the separatrix $p_{\mathrm{sep}}=0$. The value of $p_{\mathrm{foot}}$ is adjusted to meet this requirement. 

The maximum-gradient location is prescribed through the dimensionless \textsc{Varyped} control $x_{p,\mathrm{sym}}$,

\begin{equation}
\psi_{p,\mathrm{sym}}
=
1-\frac{x_{p,\mathrm{sym}}p_{\mathrm{wid}}}{2},
\label{eq:varyped-psym}
\end{equation}

and the requested overall normalized beta is constrained according to

\begin{equation}
\beta_N
=
x_{\beta_N}\beta_{N,\ell_i}\ell_i ,
\label{eq:varyped-beta}
\end{equation}

where $\ell_i$ is the internal inductance, $\beta_{N,\ell_i}$ specifies the overall-$\beta_N$ limit as a multiplier of $\ell_i$, and $x_{\beta_N}$ controls the fraction of that limit used in a particular case. For the present scan, $\beta_{N,\ell_i}=4$ and $x_{\beta_N}=0.5$ or $1.0$. The pedestal normalized-beta parameter spans $\beta_{N,\mathrm{ped}}=0.25$--$2.0$, with the precise retained discrete sequence depending on the requested $q_{95}$ and equilibrium convergence. The pedestal-width and maximum-gradient-location controls span
$p_{\mathrm{wid}}=0.04$, $0.08$, $0.12$, and $0.16$ and
$x_{p,\mathrm{sym}}=0.5$, $1.0$, and $1.5$, respectively.

\subsubsection{Current-density parameterization}

The flux-surface-averaged toroidal current-density profile is represented as the sum of a core contribution and a localized pedestal-current contribution,

\begin{equation}
j(\psi_N)
=
j_0\left(1-\psi_N^{c_1}\right)^{c_2}
+
j_{\mathrm{ped}}
\exp\left[
-\frac{(\psi_N-\psi_{j,\mathrm{sym}})^2}
       {2j_{\mathrm{wid}}^2}
\right].
\label{eq:varyped-current}
\end{equation}

Here, $j_0$, $c_1$, and $c_2$ determine the core-current profile, while $j_{\mathrm{ped}}$, $\psi_{j,\mathrm{sym}}$, and $j_{\mathrm{wid}}$ specify the amplitude, radial location, and width of the pedestal-current contribution. The profile coefficients are constrained by the prescribed values of the on-axis safety factor $q_0$, the edge safety factor $q_{95}$, the pedestal-current amplitude, and the minimum current density immediately inside the pedestal.

The pedestal-current amplitude is prescribed relative to the collisionless bootstrap-current reference used by Varyped,

\begin{equation}
j_{\mathrm{ped}}
=
x_{j,\mathrm{ped}}j_{\mathrm{ST}},
\label{eq:varyped-jped}
\end{equation}

where $j_{\mathrm{ST}}$ denotes the collisionless bootstrap-current estimate and $x_{j,\mathrm{ped}}$ is scanned over $0.5$, $0.75$, and $1.0$. The radial location of the current-density maximum is specified relative to the pressure-gradient location,

\begin{equation}
\psi_{j,\mathrm{sym}}
=
\psi_{p,\mathrm{sym}}
+
x_{j,\mathrm{sym}}p_{\mathrm{wid}},
\label{eq:varyped-jsym}
\end{equation}

with $x_{j,\mathrm{sym}}=-0.2$, $-0.1$, and $0$. Thus, the current-density maximum coincides with the pressure-gradient maximum only when $x_{j,\mathrm{sym}}=0$; the remaining values shift the current-density maximum inward relative to the pressure-gradient location.

The pedestal-current width is determined from the pressure-pedestal width and the relative displacement of the pressure- and current-profile maxima,

\begin{equation}
j_{\mathrm{wid}}
=
\frac{
p_{\mathrm{wid}}
-
\left(
\psi_{p,\mathrm{sym}}-\psi_{j,\mathrm{sym}}
\right)
}
{2^{3/2}}
+
x_{j,\mathrm{wid}},
\label{eq:varyped-jwid}
\end{equation}

where the additional width offset is fixed at
$x_{j,\mathrm{wid}}=0$ for the present database. This form for the current width reflects the fact that collisionless bootstrap current profile closely matches the pressure gradient profile while the strong increase in collisionality toward the separatrix narrows the bootstrap current profile and shifts it radially inward relative to the pressure gradient. The minimum current density immediately inside the pedestal is fixed at

\begin{equation}
j_{\min}=0.1j_{\mathrm{ped}}.
\label{eq:varyped-jmin}
\end{equation}

The on-axis safety factor is held fixed at $q_0=1.2$, while the edge safety factor is scanned over
$q_{95}=3$, $4$, $5$, $7$, and $9$. The toroidal magnetic field is fixed at $B_T=2.1$~T.

\subsubsection{Boundary shape and full profile assumptions}

The present study uses one fixed, approximately up--down-symmetric DIII-D double-null boundary shape. The shape is characterized approximately by elongation $\kappa=1.95$ and triangularity $\delta=0.8$ and is constrained to remain compatible with the DIII-D vacuum vessel and shaping-coil configuration. Small deviations from the prescribed boundary may occur during EFIT reconstruction as the pressure and current profiles are varied. Although \textsc{Varyped} can generate controlled variations of elongation, triangularity, squareness, and other boundary characteristics, plasma-shape variation is not included in the present database.

Full density and temperature profiles are generated consistently with the pressure profile so that each equilibrium can be converted into the profile and geometry inputs required by BOUT++. The pedestal density is fixed according to

\begin{equation}
\frac{n_{e,\mathrm{ped}}}{n_{\mathrm{GW}}}=0.5,
\label{eq:varyped-density}
\end{equation}

where $n_{\mathrm{GW}}$ is the Greenwald density. The density profile is represented by a hyperbolic-tangent form that is approximately flat in the core and has the same pedestal location and width as the pressure profile. The ion-to-electron temperature ratio is fixed at

\begin{equation}
\frac{T_i}{T_e}=1.4,
\label{eq:varyped-temperature}
\end{equation}

and the effective charge is assumed to be radially constant with

\begin{equation}
Z_{\mathrm{eff}}=2.2,
\label{eq:varyped-zeff}
\end{equation}

corresponding to the carbon-impurity assumption used in constructing the full profile sets. The profiles are extended through the separatrix and into the scrape-off layer and are written as profile files associated with each generated equilibrium.

The principal scanned and fixed quantities are summarized in Table~\ref{tab:varyped-parameters}. These quantities are \textsc{Varyped} input controls and profile assumptions; they should be distinguished from the equilibrium-derived physical features subsequently used as inputs to the Gaussian Process surrogate.

\begin{table*}
\caption{
Principal scanned and fixed quantities used to construct the single-shape \textsc{Varyped}/BOUT\_DB equilibrium database. The listed quantities are \textsc{Varyped} input controls and fixed profile assumptions rather than the derived physical features used for surrogate training.
}
\label{tab:varyped-parameters}
\centering
\begin{ruledtabular}
\begin{tabular}{lll}
Category & Parameter & Values or range \\
\hline
Pressure scan
& $\beta_{N,\mathrm{ped}}$
& $0.25$--$2.0$; discrete sequence depends on $q_{95}$ \\
& $p_{\mathrm{wid}}$
& $0.04,\ 0.08,\ 0.12,\ 0.16$ \\
& $x_{p,\mathrm{sym}}$
& $0.5,\ 1.0,\ 1.5$ \\
& $x_{\beta_N}$
& $0.5,\ 1.0$ \\[2pt]

Current scan
& $x_{j,\mathrm{ped}}$
& $0.5,\ 0.75,\ 1.0$ \\
& $x_{j,\mathrm{sym}}$
& $-0.2,\ -0.1,\ 0$ \\
& $q_{95}$
& $3,\ 4,\ 5,\ 7,\ 9$ \\[2pt]

Fixed equilibrium controls
& $q_0$
& $1.2$ \\
& $B_T$
& $2.1$~T \\
& $\beta_{N,\ell_i}$
& $4$ \\
& $x_{j,\mathrm{wid}}$
& $0$ \\
& $j_{\min}/j_{\mathrm{ped}}$
& $0.1$ \\
& $\kappa$
& approximately $1.95$ \\
& $\delta$
& approximately $0.8$ \\[2pt]

Fixed profile assumptions
& $n_{e,\mathrm{ped}}/n_{\mathrm{GW}}$
& $0.5$ \\
& $T_i/T_e$
& $1.4$ \\
& $Z_{\mathrm{eff}}$
& $2.2$ \\
& $p_{\mathrm{sep}}$
& $0$
\end{tabular}
\end{ruledtabular}
\end{table*}

\subsubsection{Equilibrium Convergence and Realized Database Coverage}

The structured single-shape scan contains 7,992 requested \textsc{Varyped} equilibrium configurations. Of these, 3,901 satisfy the prescribed EFIT convergence criterion, corresponding to an overall equilibrium-convergence rate of approximately 49\%. Although the requested scan systematically covers prescribed combinations of the pressure, current, and safety-factor control parameters, the resulting converged equilibrium database is not uniformly distributed in either the input-control space or the derived physical-feature space.

The convergence probability decreases for parameter combinations involving large pedestal pressure gradients, high pedestal current densities, narrow pedestal widths, and large $q_{95}$. Based on the Varyped implementation, many of these failures are associated with numerical vertical-stability difficulties encountered during EFIT reconstruction. Consequently, equilibrium-convergence filtering introduces structured gaps and under-represented regions into the realized database rather than removing cases randomly.

These convergence gaps are retained as part of the database metadata because they define the domain that is actually sampled by the subsequent \textsc{Hypnotoad} mesh-generation and \textsc{BOUT++} simulation workflows. Subsequent losses during mesh generation, \textsc{BOUT++} execution, and final quality-control filtering are comparatively small. The resulting nonuniform parameter-space coverage underscores the need to interpret the Gaussian Process posterior uncertainty relative to the successfully realized database. Within the existing candidate pool, this uncertainty can help identify comparatively under-represented regions, whereas extension into regions excluded by equilibrium-convergence failures will require targeted equilibrium generation and additional convergence studies. The 3,901 converged equilibria and their associated profile files constitute the \textsc{Varyped}/BOUT\_DB input database for the \textsc{Hypnotoad} mesh-generation stage described in the following subsection.

\subsection{Hypnotoad Mesh Generation}

Each converged \textsc{Varyped} equilibrium is converted into a field-aligned computational mesh using the \textsc{Hypnotoad} grid generator~\cite{hypnotoad_docs}. \textsc{Hypnotoad} reads the reconstructed EFIT equilibrium and generates a logically rectangular mesh aligned with magnetic flux surfaces while preserving the magnetic geometry, flux coordinates, and metric tensors required by \textsc{BOUT++}. Field-aligned meshes reduce numerical errors associated with the strong anisotropy of transport along magnetic field lines and provide the geometric quantities required by the reduced-MHD model, including the magnetic field, Jacobian, curvature operators, and metric coefficients.

For the present study, each computational mesh consists of $260\times64$ grid points in the radial ($x$) and poloidal ($y$) directions. The toroidal direction is represented spectrally within \textsc{BOUT++}, where individual toroidal harmonics are computed independently during the linear-stability calculations. Only equilibria for which \textsc{Hypnotoad} successfully generates a valid field-aligned mesh are retained for subsequent \textsc{BOUT++} simulations.

For the present single-shape database, \textsc{Hypnotoad} successfully generated computational meshes for 3,877 of the 3,901 converged \textsc{Varyped} equilibria, corresponding to a mesh-generation success rate of approximately 99.4\%. These field-aligned meshes provide the computational foundation for the \textsc{BOUT++} linear-stability calculations described in the following subsection.

\subsection{BOUT++ Linear Stability Calculations}
\label{bout}

Linear stability calculations are performed using the high-$\beta$ flute-reduced MHD model implemented in the \textsc{BOUT++} application \texttt{elm\_pb}~\cite{dudson2009bout,xu2010nonlinear,li2014linear}. The simulations are carried out in linear mode by setting \texttt{nonlinear=false}, such that the temporal evolution of small-amplitude perturbations is governed by the linearized reduced-MHD equations. The model evolves coupled perturbations of plasma vorticity, magnetic flux, and pressure, while the equilibrium pressure and current profiles are taken directly from the \textsc{Varyped}-generated equilibria.

The present ELMO database includes two completed linear physics models. The first corresponds to ideal-MHD linear-stability calculations, while the second additionally incorporates diamagnetic effects. Database generation for additional physics models, including finite resistivity and drift-Alfv\'en instability (DAI) physics, is currently underway and will be incorporated into future surrogate-model development.

For each equilibrium, individual toroidal harmonics are computed independently by retaining a single toroidal mode through the \textsc{BOUT++} toroidal filter. The toroidal mode number is scanned over

\[
n = 5\text{--}80,
\]

using an interval of

\[
\Delta n = 5,
\]

yielding sixteen toroidal mode numbers for each physics model. Linear growth rates are therefore computed for both the ideal-MHD and ideal-plus-diamagnetic models, producing two sixteen-mode growth-rate spectra (32 surrogate output quantities) for every equilibrium.

For each toroidal harmonic, the linear growth rate is extracted from the temporal evolution of the root-mean-square (RMS) pressure perturbation evaluated at the outer midplane and the radial grid point corresponding to the maximum equilibrium pressure gradient. The instantaneous logarithmic growth rate is computed as

\[
\gamma(t)=\frac{d}{dt}\ln\!\left(\mathrm{RMS}(P)\right),
\]

where $P$ denotes the pressure perturbation. The reported linear growth rate for each toroidal mode is obtained by averaging $\gamma(t)$ over the linear growth phase of the simulation. 
Growth rates are reported in normalized form as
\begin{equation}
\widehat{\gamma}
=
\frac{\gamma}{\omega_A},
\qquad
\omega_A
=
\frac{V_A}{R_0},
\qquad
V_A
=
\frac{B_0}{\sqrt{\mu_0\rho_0}},
\label{eq:growth_rate_normalization}
\end{equation}
where $\omega_A$ is the Alfv\'en frequency used in the \textsc{BOUT++} normalization, $V_A$ is the corresponding Alfv\'en velocity, $B_0$ is the reference magnetic field strength, $\rho_0$ is the reference mass density, $\mu_0$ is the vacuum permeability, and $R_0$ is the reference major radius. Accordingly, $\widehat{\gamma}$ is dimensionless, with $\widehat{\gamma}=1$ corresponding to a physical growth rate equal to $\omega_A$. All growth-rate quantities reported below, including the maximum growth rate $\widehat{\gamma}_{\max}$ and its associated root-mean-square error (RMSE) and mean absolute error (MAE), use this normalized convention.

The resulting values are assembled into separate sixteen-mode growth-rate spectra for the ideal-MHD and ideal-plus-diamagnetic physics models. For each equilibrium $i$ and physics model $m\in\{\mathrm{ideal},\mathrm{dia}\}$, the maximum growth rate is derived from the corresponding sixteen-mode spectrum as 
\begin{equation}
\widehat{\gamma}_{\max,i}^{(m)}
=
\max_{k=1,\ldots,16}
\widehat{\gamma}_i^{(m)}(n_k),
\qquad
n_k=5,10,\ldots,80.
\label{eq:gmax_by_physics}
\end{equation}
The corresponding dominant toroidal mode is
\begin{equation}
n_{\max,i}^{(m)}
=
\underset{n_k}{\operatorname{arg\,max}}\,
\widehat{\gamma}_i^{(m)}(n_k).
\label{eq:nmax_by_physics}
\end{equation}

The two sixteen-mode spectra constitute the 32 Gaussian Process target quantities. The quantities $\widehat{\gamma}_{\max}$ and $n_{\max}$ are derived by post-processing the spectra; they are not trained as separate surrogate outputs. Mode-structure diagnostics and pipeline quality-control flags are retained as database metadata but are likewise not direct outputs of the present Gaussian Process model.

The same post-processing definitions are applied separately to the sixteen surrogate-predicted outputs for each physics model to obtain $\widehat{\gamma}_{\max,{\mathrm{GP}},i}^{(m)}$ and $n_{\max,{\mathrm{GP}},i}^{(m)}$. Consequently, performance in predicting the maximum growth rate and dominant toroidal mode is assessed separately for the ideal-MHD and ideal-plus-diamagnetic models.

\subsection{Automated Campaign Execution and Reduced-Data Extraction}

The \textsc{BOUT++} simulation database comprises independent calculations spanning thousands of equilibria, two linear physics models, and sixteen toroidal mode numbers. Python-based workflow composition with EFFIS~\cite{effis} was used to prepare run directories, apply physics and toroidal-mode settings, launch and monitor jobs, identify incomplete or failed calculations, and restart cases requiring additional integration time.

Completed simulations were automatically post-processed in parallel to extract the linear growth rates defined in Section~\ref{bout} from the temporal evolution of the RMS pressure perturbation. Mode-resolved results were then aggregated for each equilibrium to construct the ideal-MHD and ideal-plus-diamagnetic growth-rate spectra. The final reduced dataset includes these spectra, the maximum linear growth rate ($\widehat{\gamma}_{\max}$), the corresponding dominant unstable toroidal mode ($n_{\max}$), pipeline quality-control flags, representative mode-structure diagnostic visualizations, and the associated equilibrium-derived pedestal features. This reduction decreases the data volume from approximately 11~MB to 0.3~MB per stored time step while retaining the quantities required for stability analysis and surrogate-model development.

Both the raw simulation data and the post-processed analysis products are stored using ADIOS~\cite{ADIOS2}. Campaign metadata\footnote{
Campaign management with scientific data using ADIOS, HDF5, text, images, etc. is an emerging topic, with an overview publication in preparation. See \url{https://github.com/ornladios/hpc-campaign/tree/master} for the software referenced here.
} preserve the associations among each equilibrium, computational mesh, \textsc{BOUT++} configuration, execution status, and reduced analysis products. Broader campaign indexing, automated provenance tracking, and surrogate-triggered execution of new high-fidelity simulations remain future capabilities of the ELMO framework.

The simulation and workflow calculations reported here used \textsc{BOUT++}, \textsc{Hypnotoad}, EFFIS, ADIOS, and the HPC Campaign software versions listed in table II. These version identifiers are recorded together with the campaign metadata to support reproducibility of the simulation and data-reduction workflow.

\subsection{Simulation Database and Quality Control}
\label{database}

The ELMO simulation database is constructed by combining systematic equilibrium generation using \textsc{Varyped}, automated field-aligned mesh generation with \textsc{Hypnotoad}, and large-scale linear-stability calculations performed using \textsc{BOUT++}. Case identifiers and campaign metadata preserve the associations among equilibrium parameters, reconstructed equilibria, computational meshes, \textsc{BOUT++} configurations, execution status, and reduced stability outputs. This organized simulation database provides the physics foundation for subsequent surrogate-model development.

For the present single-shape study, 7,992 equilibrium configurations were generated using \textsc{Varyped}, of which 3,901 converged successfully during EFIT reconstruction. \textsc{Hypnotoad} subsequently generated valid field-aligned meshes for 3,877 equilibria. Of these, 3,869 completed the subsequent \textsc{BOUT++} linear-stability calculations successfully without numerical issues and were retained for surrogate-model development. The excluded cases correspond to simulations that did not complete successfully or exhibited numerical issues during post-processing and therefore did not satisfy the quality-control criteria.

For each retained equilibrium, linear growth rates were computed for sixteen toroidal mode numbers ($n=5$--80 with $\Delta n=5$) using both the ideal-MHD and ideal-plus-diamagnetic physics models, yielding two sixteen-mode growth-rate spectra (32 surrogate output quantities) per equilibrium. The resulting database therefore contains

\[
3,\!869 \times 32 = 123,\!808
\]

mode-resolved \textsc{BOUT++} linear-stability calculations.

For each equilibrium, the reduced database stores the two growth-rate spectra together with the maximum linear growth rate ($\widehat{\gamma}_{\max}$), the dominant unstable toroidal mode ($n_{\max}$), stability classification, representative mode-structure diagnostics, and the corresponding equilibrium-derived pedestal features. The characteristics of the resulting quality-controlled database are described in Section~\ref{sec:statistics}, while its use for surrogate-model development is presented in Section~\ref{sec:gpr}.

\section{Uncertainty-aware Gaussian Process Surrogate Model}
\label{sec:workflow_ml}

\begin{figure*}[t]
  \centering
  \begin{tikzpicture}[
    box/.style={
      rectangle,
      rounded corners=3pt,
      draw=black,
      thick,
      align=center,
      minimum width=2cm,
      minimum height=0.9cm,
      font=\small
    },
    simulation/.style={
      box,
      draw=orange!90!black,
      fill=orange!5
    },
    finalbox/.style={
      box,
      draw=green!55!black,
      fill=green!5
    },
    decision/.style={
      diamond,
      aspect=2.4,
      draw=black,
      thick,
      align=center,
      text width=2.9cm,
      inner sep=1.5pt,
      font=\small
    },
    looparrow/.style={
      -{Latex[length=3mm]},
      ultra thick,
      draw=violet!90!black
    },
  ]

  \node[box, fill=gray!5] (init) {Initial\\training set};
  \node[box, fill=gray!5, right=0.7cm of init] (train) {Train GP\\surrogate};
  
  \node[box, draw=violet!90!black, fill=violet!5, right=0.7cm of train, text width=2.33cm] (predict) {Predict $\mu_j(\mathbf{x})$ and $\sigma_{f,j}(\mathbf{x})$};
  
  \node[box, draw=violet!90!black, fill=violet!5, right=1cm of predict] (acq) { Acquisition function\\$U(\mathbf{x})=\frac{1}{N_{\rm out}}\sum_j\sigma_{f,j}$};
  
  \node[box, draw=violet!90!black, fill=violet!5, right=1cm of acq, text width=2.9cm] (select) {Select largest-\\acquisition samples};
  
  \node[simulation, below=0.5cm of select, xshift=2.5cm] (simulate){New Run high-\\fidelity simulations};
  
  \node[box, draw=violet!90!black, fill=violet!5, below=1.8cm of select, text width=2.7cm] (update) {Update training\\ dataset};
  
  \node[box, draw=violet!90!black, fill=violet!5, left=1cm of update] (retrain){Retrain GP\\ surrogate};
  
  \node[decision, draw=violet!90!black, fill=violet!5, left=1cm of retrain] (stop)
  {Held-out validation\\[1mm]
  $\Delta R^2<5\times10^{-4}$};
  
  \node[finalbox, left=1cm of stop] (final) {Final GP surrogate \& \\Independent test \\ evaluation };
  
  \draw[arrow] (init) -- (train);
  \draw[arrow] (train) -- (predict);
  \draw[looparrow] (predict) -- (acq);
  \draw[looparrow] (acq) -- (select);
  \draw[looparrow] (select) -- (update);
  
  \draw[dash_arrow, orange!90!black] (select.east) -- ++(0.5,0)  -| (simulate.north);
  \draw[dash_arrow, orange!90!black] (simulate.south) -- (simulate.south |- update.east) -- (update.east);
  
  \draw[looparrow] (update) -- (retrain);
  \draw[looparrow] (retrain) -- (stop);
  \draw[arrow] (stop.west) -- node[above]{Yes} (final.east);
  \draw[looparrow] (stop.north) -- node[right]{No} (predict.south);


  \end{tikzpicture}
  \caption{\textbf{Adaptive Gaussian-process workflow.} Starting from an initial training set, a Gaussian-process (GP) surrogate is constructed and used to predict the posterior mean, $\mu_j(\mathbf{x})$, and latent-function posterior standard deviation, $\sigma_{f,j}(\mathbf{x})$, for each of the 32 output channels of every candidate equilibrium in the adaptive-training pool. The acquisition function averages the standardized posterior standard deviations across the output channels, and the candidates having the largest acquisition scores are selected for inclusion in the training set. The GP surrogate is then retrained using the expanded training set, and its validation performance is evaluated on the fixed validation set. If the prescribed stopping criterion is satisfied, model development is terminated and the performance of the frozen final surrogate is reported on the independent test set; otherwise, the adaptive-learning cycle is repeated. The independent test set is not used for feature selection, adaptive sample acquisition, learning-curve construction, or stopping. The current pool-based adaptive-learning workflow is highlighted by the violet boxes and arrows. The orange dashed pathway illustrates a future closed-loop extension in which newly generated high-fidelity simulations are incorporated to expand the training database iteratively.}

  \label{fig:adaptive_gp_loop}
\end{figure*}
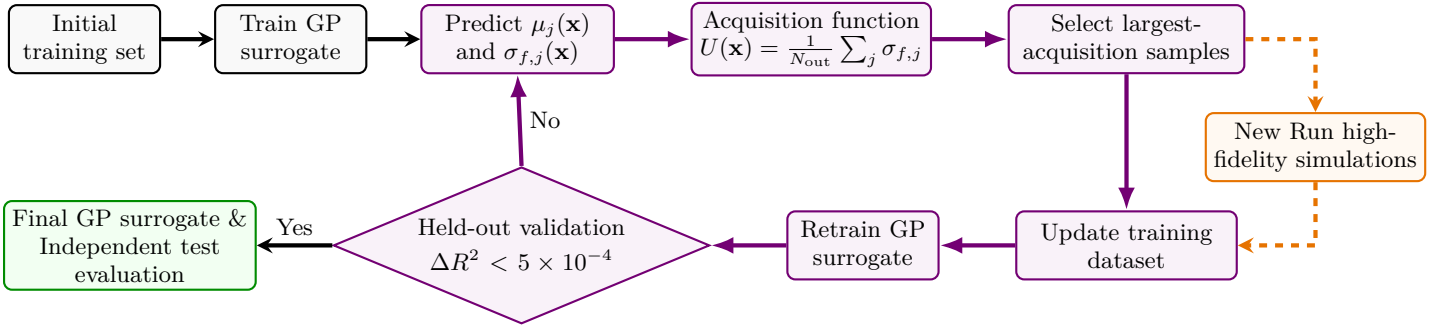

The surrogate model maps equilibrium-derived pedestal features to the corresponding \textsc{BOUT++} linear growth-rate spectra, providing a computationally efficient approximation to the high-fidelity simulations while also estimating latent posterior uncertainty. For each equilibrium, the surrogate predicts the two sixteen-mode growth-rate spectra corresponding to the ideal-MHD and ideal-plus-diamagnetic physics models. The posterior uncertainty is subsequently used as a relative indicator of surrogate uncertainty and to guide pool-based adaptive sample selection.

Gaussian Process Regression (GPR)~\cite{michoski2024gaussian} is adopted because it provides both a posterior mean and a latent-function posterior standard deviation through a Bayesian probabilistic formulation~\cite{rasmussen2006gaussian}. These characteristics make GPR well suited for surrogate modeling of computationally expensive simulations and for uncertainty-guided adaptive sample selection.

Figure~\ref{fig:adaptive_gp_loop} summarizes the surrogate-training workflow. Starting from an initial randomly selected training subset drawn from the adaptive-training pool, the Gaussian Process surrogate is trained using the selected pedestal features. Posterior uncertainty is then evaluated for the remaining candidate equilibria in the adaptive-training pool, which are ranked according to the acquisition function and iteratively incorporated into the training set. After each adaptive iteration, surrogate performance is evaluated using the fixed held-out validation set until the prescribed stopping criterion is satisfied. In the present study, this adaptive procedure is demonstrated entirely within the existing simulation database to evaluate the effectiveness of pool-based uncertainty-guided sample selection. Extension to a fully closed-loop workflow, in which new \textsc{Varyped} equilibria, \textsc{Hypnotoad} meshes, and \textsc{BOUT++} simulations are generated automatically according to the surrogate uncertainty, is reserved for future work.

\subsection{Gaussian Process Regression}
\label{sec:gpr}

The surrogate is implemented using the \texttt{GPyTorch} framework~\cite{gardner2018gpytorch} together with the BoTorch probabilistic modeling library~\cite{balandat2020botorch}. 
The software environment used for surrogate training and evaluation is summarized in Table~II.
The 32 surrogate output quantities are modeled using independent single-task Gaussian Processes, with one Gaussian Process assigned to each output channel. For each equilibrium, the surrogate predicts the two sixteen-mode growth-rate spectra corresponding to the ideal-MHD and ideal-plus-diamagnetic physics models.

For a training dataset
\begin{equation}
\mathcal{D}
=
\left\{
(\mathbf{x}_i,\mathbf{y}_i)
\right\}_{i=1}^{N},
\end{equation}
where $\mathbf{x}_i$ denotes the vector of input features and $\mathbf{y}_i$ denotes the corresponding vector of 32 mode-resolved growth-rate outputs, the Gaussian Process prior for output channel $j$ is
\begin{equation}
f_j(\mathbf{x})
\sim
\mathcal{GP}
\left(
\mu_j(\mathbf{x}),
k_j(\mathbf{x},\mathbf{x}')
\right),
\end{equation}
where $\mu_j(\mathbf{x})$ and $k_j(\mathbf{x},\mathbf{x}')$ are the mean function and covariance kernel, respectively, for output channel $j$. A learnable constant mean function is adopted together with an isotropic Matérn covariance kernel having smoothness parameter $\nu=2.5$. Accordingly, each covariance model uses a common characteristic length scale across the input dimensions. The model hyperparameters, including the constant mean, characteristic length scale, output scale, and Gaussian-likelihood noise, are determined by maximizing the exact marginal likelihood. The 32 mode-resolved output quantities are modeled as conditionally independent Gaussian Process outputs.

Given the training data for output channel $j$, the latent-function posterior distribution at a new equilibrium $\mathbf{x}^{*}$ is
\begin{equation}
f_j(\mathbf{x}^{*})\mid\mathcal{D}_j
\sim
\mathcal{N}
\left(
\mu_j(\mathbf{x}^{*}),
\sigma_{f,j}^{2}(\mathbf{x}^{*})
\right),
\end{equation}
where $\mu_j(\mathbf{x}^{*})$ is the posterior-mean prediction and $\sigma_{f,j}(\mathbf{x}^{*})$ is the corresponding latent-function posterior standard deviation. The posterior mean provides the surrogate prediction, while the posterior standard deviation provides a model-based estimate of latent surrogate uncertainty. The reported value of $\sigma_{f,j}$ excludes the additional Gaussian-likelihood observation-noise contribution and therefore should not be interpreted as a fully calibrated predictive standard deviation.

Exact Gaussian Process inference is performed using the Blackbox Matrix--Matrix (BBMM) algorithms implemented in \texttt{GPyTorch}~\cite{gardner2018gpytorch}, enabling efficient training and prediction for the moderate-sized database considered in this study. Input features are standardized independently to zero mean and unit variance, while the ideal-MHD and ideal-plus-diamagnetic output blocks are standardized separately before training. The corresponding normalization statistics are calculated using only the 70\% adaptive-training pool for each data-partition realization and are then applied unchanged to the validation and independent test subsets. Following prediction, all output quantities and posterior standard deviations are transformed back into the original \textsc{BOUT++} normalized representation for subsequent performance evaluation and comparison with the corresponding \textsc{BOUT++} calculations.

The resulting model-based posterior uncertainty is subsequently used in the pool-based adaptive sample-selection procedure described in Section~\ref{sec:UQ} to rank candidate equilibria for inclusion in the adaptive-training set.

\subsection{Physics-informed Feature Selection}
\label{sec:feature}

\begin{figure*}[t]
    \centering
    \begin{tikzpicture}[node distance=1.6cm, font=\footnotesize,
        box/.style={
        rectangle,
        rounded corners,
        draw=black,
        align=center,
        minimum height=1.0cm,
        fill=gray!10},]
        
        \node (pro1) 
        [process, text width=3.5cm]
        {\textbf{Train GP surrogate}\\ 10 features
        \vspace{0.1cm}
        };
        
        \node (pro2) 
        [process, right of=pro1, text width=5cm, xshift=4cm,]
        {\textbf{Compute feature-selection diagnostics} \\
        \vspace{0.15cm}
        $\bullet$ Permutation importance {$\Delta$ RMSE}\\
        \vspace{0.1cm}
        $\bullet$ Spearman correlation\\
        \vspace{0.1cm}
        $\bullet$ Variance inflation factor (VIF)
        };
        
        \node (pro3) 
        [process, right of=pro2, text width=4.1cm, 
        xshift=4cm, ] 
        {\textbf{Identify candidate features}\\
        \vspace{0.15cm}
        $\bullet$ Low permutation importance\\
        \vspace{0.1cm}
        $\bullet$ High Spearman correlation\\
        \vspace{0.1cm}
        $\bullet$ High VIF
        };
        
        \node (pro4) 
        [process, below of=pro3, 
        text width=4cm,
        yshift=-1.5cm] 
        {\textbf{Remove candidate feature and retrain GP}
        \vspace{0.1cm}
        };
        
        \draw [arrow] (pro1) -- (pro2);
        \draw [arrow] (pro2) -- (pro3);
        \draw [arrow] (pro3) -- (pro4);
        
        \node (dec1) 
        [decision, left of=pro4, 
        minimum width=2cm, 
        minimum height=1.5cm,
        text width=2.3cm,
        xshift=-4.cm
        ] 
        {{$\Delta R^2 < 0.005$?}
        };
        \draw [arrow] (pro4) -- (dec1);

        \node (stop1) 
        [startstop, below of=dec1, text width=2.45cm, 
        yshift=-0.5cm] 
        {\textbf{Stop} \\ Final feature set.};
        \draw [arrow] (dec1) -- node[anchor=west] {No} (stop1);

        \node (pro5) 
        [process, left of=dec1,
        minimum width=2cm, 
        minimum height=1.5cm,
        text width=3.5cm,
        xshift=-4.cm,] 
        {\textbf{Accept removal} \\ Update feature set.
        };
        \draw [arrow] (dec1) -- node[anchor=south] {Yes} (pro5);
        \draw [arrow] (pro5) -- (pro2);
        
    \end{tikzpicture}
    \caption{\textbf{Physics-informed recursive feature-elimination workflow used to identify the final Gaussian Process input feature set.} Starting from the original ten-feature model, permutation feature importance, Spearman rank correlation, and variance inflation factor (VIF) analyses are used to identify candidate removable features. Each candidate feature is removed individually, the Gaussian Process surrogate is retrained using samples drawn exclusively from the 70\% adaptive-training pool, and its predictive performance is evaluated on the fixed 15\% validation set. The independent 15\% test set is not accessed during feature selection, Gaussian Process fitting, adaptive sample acquisition, or stopping. A feature is permanently removed only if the reduction in the mean coefficient of determination satisfies the acceptance criterion, $\Delta R^2<0.005$. After each accepted removal, the feature-selection diagnostics are recomputed using the updated feature set, and the procedure is repeated until no additional feature satisfies the acceptance criterion. The final model retains eight physically meaningful input features while maintaining essentially the same predictive accuracy as the original ten-feature model.}
    \label{fig:feature_selection_workflow}
\end{figure*}
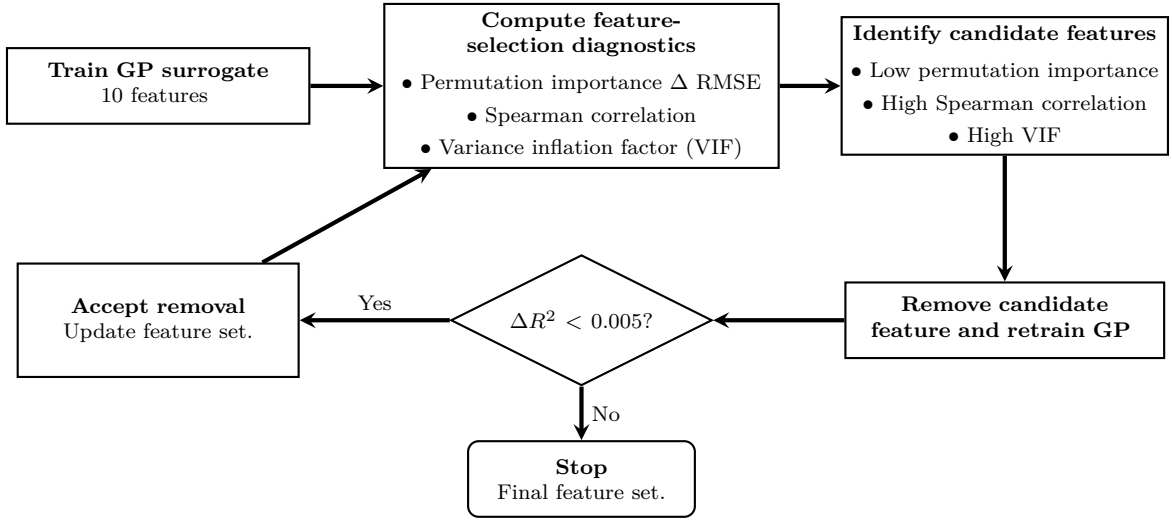

Rather than using the complete equilibrium profiles as surrogate inputs, the Gaussian Process model is constructed from a compact set of equilibrium-derived scalar features describing the pedestal pressure, current density, and magnetic equilibrium. This reduced representation lowers the input dimensionality, improves computational efficiency, and preserves physical interpretability.

For each converged \textsc{BOUT++} simulation, the \textsc{Varyped}-generated equilibrium profiles are post-processed to extract an initial set of ten physically motivated features:

\begin{enumerate}
\item peak normalized pressure-gradient parameter, $\alpha_{\rm peak}$,
\item maximum pressure gradient, $(dp/dr)_{\rm max}$,
\item peak current density, $j_{\rm peak}$,
\item radial location of $\alpha_{\rm peak}$,
\item radial location of $(dp/dr)_{\rm max}$,
\item radial location of $j_{\rm peak}$,
\item pedestal center, $\psi_{\rm ped}$,
\item pedestal width, $\Delta_{\rm ped}$,
\item pedestal height, $p_{\rm ped}$,
\item edge safety factor, $q_{95}$.
\end{enumerate}

These quantities provide a compact representation of the equilibrium characteristics governing peeling--ballooning stability. The corresponding feature vector is

\begin{equation}
\mathbf{x} =
\left[
\begin{aligned}
&\alpha_{\rm peak},
(dp/dr)_{\rm max},
j_{\rm peak},
\psi_{\alpha},
\psi_{dp/dr},\\
&\psi_{j},
\psi_{\rm ped},
\Delta_{\rm ped},
p_{\rm ped},
q_{95}
\end{aligned}
\right]^{\mathrm T}.
\end{equation}

For each equilibrium, the surrogate predicts the two sixteen-mode growth-rate spectra computed by \textsc{BOUT++} for the ideal-MHD and ideal-plus-diamagnetic physics models,

\begin{equation}
\mathbf{y}
=
\left[
\widehat\gamma_{\rm ideal}(n_1),
\ldots,
\widehat\gamma_{\rm ideal}(n_{16}),
\widehat\gamma_{\rm dia}(n_1),
\ldots,
\widehat\gamma_{\rm dia}(n_{16})
\right]^{\mathrm T},
\end{equation}

yielding 32 output quantities for every equilibrium.

Prior to training, each input feature is standardized independently to zero mean and unit variance,

\begin{equation}
x_i' = \frac{x_i-\mu_i}{\sigma_i},
\end{equation}

where $\mu_i$ and $\sigma_i$ denote the mean and standard deviation of the $i$th feature. The ideal-MHD and ideal-plus-diamagnetic output channels are standardized separately before Gaussian Process training and transformed back into the original BOUT++ normalized representation for subsequent performance evaluation.

Feature importance is first evaluated using permutation feature importance. Each feature is randomly permuted while all remaining features are held fixed, and the resulting increase in prediction error is used as a measure of feature importance. This model-agnostic approach provides an intuitive assessment of the relative contribution of each feature to surrogate performance.

The final feature set is determined using a physics-informed recursive backward feature-elimination procedure (Fig.~\ref{fig:feature_selection_workflow}). Rather than relying on a single statistical criterion, candidate features are identified using four complementary diagnostics:

\begin{enumerate}
\item permutation feature importance,
\item Spearman rank correlation,
\item variance inflation factor (VIF), and
\item surrogate retraining with performance evaluation.
\end{enumerate}

Candidate selection considers the combined evidence from permutation importance, feature correlation, multicollinearity, and surrogate performance rather than relying on any single diagnostic, allowing correlated features with redundant predictive information to be removed while preserving overall model accuracy.

At each iteration, each candidate feature is removed individually, the complete adaptive Gaussian Process workflow is retrained using the reduced feature set, and surrogate performance is evaluated exclusively on the fixed validation set five random data-partition realizations. The candidate producing the highest mean predictive accuracy is accepted only if the corresponding reduction in the mean coefficient of determination remains below the prescribed acceptance threshold. The independent test set is not accessed during feature selection. The feature-selection diagnostics are then recomputed using the updated feature set, and the procedure is repeated until no additional feature satisfies the removal criterion.

The resulting feature rankings and the performance of the recursive feature-elimination procedure are presented in Section~\ref{sec:verification}.

\subsection{Surrogate Training, Validation, and Independent Testing}
\label{sec:train}

The simulation database is partitioned at the equilibrium level, with all 32 mode-resolved outputs from a given equilibrium retained in the same subset. For each realization, approximately 70\% of the 3,869 equilibria are assigned to the adaptive-training pool, 15\% to a fixed validation set, and the remaining 15\% to an independent test set. All initial sampling, uncertainty-guided candidate selection, and Gaussian Process fitting are performed exclusively within the adaptive-training pool. The validation set remains fixed throughout model development and is used for feature-selection decisions, learning-curve monitoring, and determination of the adaptive stopping point. The independent test set is completely excluded from Gaussian Process fitting, feature selection, sample acquisition, and stopping decisions and is evaluated only after the final feature set and training procedure have been fixed.

\begin{figure*}[t]
  \centering
  \includegraphics[width=0.95\textwidth]{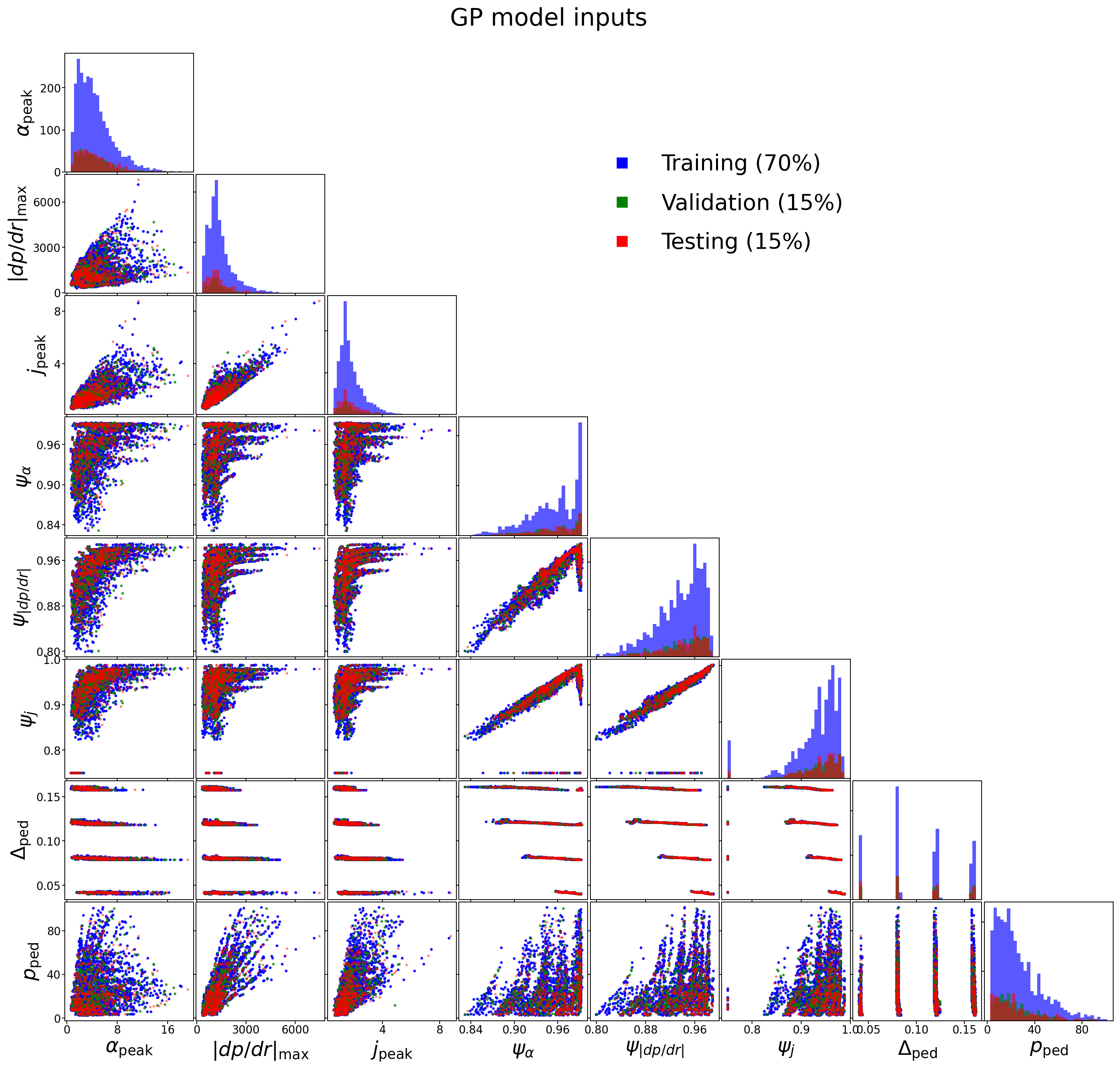}
  \caption{\textbf{Distributions and pairwise correlations of the eight input features used for the final Gaussian Process surrogate.} The diagonal panels show the marginal distributions of each input feature, while the off-diagonal panels show their pairwise relationships. Blue, green, and red denote the adaptive-training pool (70\%), validation set (15\%), and independent test set (15\%), respectively. The three subsets exhibit closely comparable marginal distributions and pairwise relationships, indicating representative coverage of the quality-controlled simulation database. The validation set is used for model-development decisions, whereas the test set remains excluded from model development and is reserved for final performance evaluation.}
  \label{fig:gp_inputs}
\end{figure*}

Figure~\ref{fig:gp_inputs} compares the distributions and pairwise relationships of the eight input features retained in the final surrogate model for the adaptive-training, validation, and independent test subsets. The three subsets exhibit closely comparable marginal distributions and pairwise relationships, indicating that both the validation and test subsets provide representative coverage of the parameter domain sampled by the quality-controlled simulation database.

The surrogate is implemented as 32 conditionally independent single-task Gaussian Process outputs, one for each mode-resolved target quantity. The model employs the isotropic Matérn covariance formulation described in Section~\ref{sec:gpr}, with smoothness parameter $\nu=2.5$. For each output GP, the isotropic kernel uses one characteristic length scale shared across the input dimensions; the kernel hyperparameters are optimized independently for the different output channels. Model hyperparameters are determined by maximizing the exact marginal likelihood using the \texttt{GPyTorch}/BoTorch implementation.

\subsubsection*{Evaluation Metrics}

Surrogate performance is quantified using the coefficient of determination ($R^2$), root-mean-square error (RMSE), and mean absolute error (MAE). During adaptive training, these metrics are evaluated on the fixed validation subset after each adaptive-training iteration and are used for model-development decisions, including assessment of convergence. The independent test subset is evaluated separately and is not used to guide adaptive sampling or model selection.

Let $y_{ij}$ and ${y}_{\mathrm{GP},ij}$ denote the BOUT++ and surrogate-predicted growth rates, respectively, for equilibrium $i$ and output channel $j$. The combined surrogate contains 32 output channels: 16 toroidal mode numbers for the ideal-MHD model and 16 corresponding channels for the ideal-plus-diamagnetic model. For each output channel, the coefficient of determination is
\begin{equation}
R_j^2
=
1-
\frac{
\displaystyle\sum_{i=1}^{N}
\left(y_{ij}-{y}_{\mathrm{GP},ij}\right)^2
}{
\displaystyle\sum_{i=1}^{N}
\left(y_{ij}-\overline{y}_j\right)^2
},
\label{eq:r2_channel}
\end{equation}
where $\overline{y}_j$ is the mean \textsc{BOUT++} growth rate for channel $j$ over the evaluation subset. The reported mean coefficient of determination is the unweighted average over all 32 output channels, 
\begin{equation}
\overline{R^2}
=
\frac{1}{32}
\sum_{j=1}^{32} R_j^2.
\label{eq:r2_mean}
\end{equation}
Thus, the sixteen ideal-MHD and sixteen ideal-plus-diamagnetic output channels contribute equally to the reported learning-curve score.

The RMSE and MAE used in the adaptive-training learning curves are likewise first calculated separately for each output channel,
\begin{equation}
\mathrm{RMSE}_j
=
\left[
\frac{1}{N}
\sum_{i=1}^{N}
\left(
{y}_{\mathrm{GP},ij}-y_{ij}
\right)^2
\right]^{1/2},
\label{eq:rmse_channel}
\end{equation}
and
\begin{equation}
\mathrm{MAE}_j
=
\frac{1}{N}
\sum_{i=1}^{N}
\left|
{y}_{\mathrm{GP},ij}-y_{ij}
\right|.
\label{eq:mae_channel}
\end{equation}
These quantities are then averaged over the 32 output channels,
\begin{equation}
\overline{\mathrm{RMSE}}
=
\frac{1}{32}
\sum_{j=1}^{32}
\mathrm{RMSE}_j,
\qquad
\overline{\mathrm{MAE}}
=
\frac{1}{32}
\sum_{j=1}^{32}
\mathrm{MAE}_j.
\label{eq:mean_error_metrics}
\end{equation}
These channel-averaged quantities, together with $\overline{R^2}$, are the metrics reported in the adaptive-training learning curves.

In addition to the full growth-rate spectrum, the predicted maximum linear growth rate, $\widehat{\gamma}_{\max}$, and the corresponding dominant unstable toroidal mode number, $n_{\max}$, are evaluated separately for the ideal-MHD and ideal-plus-diamagnetic models. For physics model $m$, the true and predicted maximum growth rates for equilibrium $i$ are obtained from the corresponding 16-channel spectra as
\begin{equation}
\widehat{\gamma}_{\max,i}^{(m)}
=
\max_{n}\widehat{\gamma}_i^{(m)}(n),
\qquad
\widehat{\gamma}_{\max,\mathrm{GP},i}^{(m)}
=
\max_{n}\widehat{\gamma}_{\mathrm{GP},i}^{(m)}(n).
\label{eq:gmax_definition}
\end{equation}
The corresponding true and predicted dominant mode numbers are
\begin{equation}
n_{\max,i}^{(m)}
=
\underset{n}{\operatorname{arg\,max}}\,
\widehat{\gamma}_i^{(m)}(n),
\qquad
{n}_{\max,\mathrm{GP},i}^{(m)}
=
\underset{n}{\operatorname{arg\,max}}\,
\widehat{\gamma}_{\mathrm{GP},i}^{(m)}(n).
\label{eq:nmax_definition}
\end{equation}

For $\widehat\gamma_{\max}^{(m)}$, the RMSE and MAE are calculated directly over the
$N$ equilibria in the evaluation subset,
\begin{equation}
\mathrm{RMSE}_{\gamma_{\max}}^{(m)}
=
\left[
\frac{1}{N}
\sum_{i=1}^{N}
\left(
\widehat{\gamma}_{\max,\mathrm{GP},i}^{(m)}
-
\widehat\gamma_{\max,i}^{(m)}
\right)^2
\right]^{1/2},
\label{eq:rmse_gmax}
\end{equation}
and
\begin{equation}
\mathrm{MAE}_{\gamma_{\max}}^{(m)}
=
\frac{1}{N}
\sum_{i=1}^{N}
\left|
\widehat{\gamma}_{\max,\mathrm{GP},i}^{(m)}
-
\widehat\gamma_{\max,i}^{(m)}
\right|.
\label{eq:mae_gmax}
\end{equation}
The corresponding mean relative error is
\begin{equation}
\mathrm{MRE}_{\gamma_{\max}}^{(m)}
=
\frac{100\%}{N}
\sum_{i=1}^{N}
\frac{
\left|
\widehat{\gamma}_{\max,\mathrm{GP},i}^{(m)}
-
\widehat\gamma_{\max,i}^{(m)}
\right|
}{
\max\left(
\left|\widehat\gamma_{\max,i}^{(m)}\right|,
\epsilon
\right)
},
\label{eq:mre_gmax}
\end{equation}
where $\epsilon=10^{-12}$ is a small numerical floor used to avoid division by zero in the relative-error calculation.

The coefficient of determination for the maximum growth rate is calculated directly over the $N$ equilibrium-wise values,
\begin{equation}
R_{\gamma_{\max}}^{2,(m)}
=
1-
\frac{
\displaystyle\sum_{i=1}^{N}
\left(
\widehat\gamma_{\max,i}^{(m)}
-
\widehat{\gamma}_{\max,\mathrm{GP},i}^{(m)}
\right)^2
}{
\displaystyle\sum_{i=1}^{N}
\left(
\widehat\gamma_{\max,i}^{(m)}
-
\overline{\widehat\gamma}_{\max}^{(m)}
\right)^2
},
\label{eq:r2_gmax}
\end{equation}
where $\overline{\widehat\gamma}_{\max}^{(m)}$ denotes the mean BOUT++ maximum growth rate for physics model $m$ over the evaluation subset. Consequently, the $R^2$ values reported for $\widehat\gamma_{\max}$ are calculated directly from the maximum-growth-rate values and are distinct from the 32-channel $\overline{R^2}$ used in the adaptive-training learning curves.

The dominant-mode prediction $n_{\max}$ is obtained independently from the BOUT++ and surrogate-predicted growth-rate spectra. Its signed prediction error for equilibrium $i$ and physics model $m$ is defined as
\begin{equation}
\Delta n_{\max,i}^{(m)}
=
{n}_{\max,\mathrm{GP},i}^{(m)}
-
n_{\max,i}^{(m)}.
\label{eq:nmax_error}
\end{equation}
Performance in predicting $n_{\max}$ is further characterized by the exact prediction fraction, the mean absolute error in $n_{\max}$, and the fractions of predictions lying within one and two sampled mode-number intervals of the \textsc{BOUT++} value.


To reduce sensitivity to the initial random partition and training subset, the complete adaptive-training procedure was repeated using five random-seed values: 123, 456, 789, 1024, and 2025. Each seed defined an independently generated partition into a 70\% adaptive-training pool, a fixed 15\% validation subset, and a fixed 15\% test subset. The subsets are mutually exclusive within each realization, although subsets generated for different realizations may overlap. Learning curves and other model-development results are reported as the mean over the five realizations, with shaded regions indicating one sample standard deviation.

For each realization, the validation subset is used for model selection and convergence assessment, whereas the corresponding fixed test subset is not used to guide adaptive sampling, feature selection, stopping, or model selection. After the training procedure and feature set have been fixed, the final generalization metrics are calculated separately on the test subset of each realization and are then summarized by their mean and sample standard deviation over the five realizations.

\subsection{Uncertainty Quantification and Pool-Based Adaptive Selection}
\label{sec:UQ}
A principal advantage of Gaussian Process Regression is that it naturally provides a latent posterior uncertainty estimate together with the surrogate prediction. For each input feature vector $\mathbf{x}$ and output channel $j$, the surrogate returns a posterior mean, $\mu_j(\mathbf{x})$, and a latent-function posterior standard deviation, $\sigma_{f,j}(\mathbf{x})$. The posterior mean represents the predicted mode-resolved growth rate, while $\sigma_{f,j}(\mathbf{x})$ provides a model-based estimate of surrogate uncertainty for that output channel. Regions located near previously observed training samples generally exhibit smaller posterior uncertainty, whereas sparsely sampled regions of parameter space generally produce larger posterior uncertainty.

In the present study, the latent posterior uncertainty is used as a relative acquisition score to guide pool-based adaptive sample selection within the adaptive-training pool. Candidate equilibria are ranked according to the acquisition function
\begin{equation}
U(\mathbf{x})
=
\frac{1}{N_{\rm out}}
\sum_{{j}=1}^{N_{\rm out}}
{\sigma_{f,j}}(\mathbf{x}),
\label{eq:acquisition}
\end{equation}
where $N_{\rm out}=32$ denotes the number of surrogate output quantities and ${\sigma_{f,j}}(\mathbf{x})$ is the latent posterior standard deviation of the $j$th output channel. The standard deviations entering Eq.~\eqref{eq:acquisition} are evaluated in the standardized output space so that the different output channels contribute on comparable scales. Candidate equilibria with the largest acquisition values are selected from the adaptive-training pool and incorporated into the training set, after which the surrogate is retrained. This strategy preferentially samples candidate equilibria having the largest mean latent posterior uncertainty across the 32 output channels while reducing repeated sampling in regions already well represented by the training data.

Surrogate performance is evaluated after each adaptive iteration using the fixed validation set described in Section~\ref{sec:train}. Adaptive training begins with $N_0=200$ equilibria randomly selected from the 70\% adaptive-training pool. At each iteration, $N_b=200$ candidate equilibria having the largest acquisition values are added to the training set, after which the Gaussian Process surrogate is retrained and evaluated on the fixed validation set.

The signed incremental improvement in validation performance at iteration $k$ is defined as
\begin{equation}
\Delta\overline{R^2}_k
=
\overline{R^2}_k-\overline{R^2}_{k-1},
\label{eq:delta_r2_stop}
\end{equation}
where $\overline{R^2}_k$ is the mean of the 32 channel-wise validation $R^2$ values after iteration $k$. Adaptive training is terminated when
\begin{equation}
\Delta\overline{R^2}_k < 5\times10^{-4}.
\label{eq:r2_stop_threshold}
\end{equation}
The stopping criterion is applied upon its first occurrence. Training also terminates when all candidate equilibria have been incorporated into the training set or when a prescribed maximum training-set size is reached, if one is specified. In the present calculations, no additional maximum training-set size was specified. The independent test set is not used in evaluating the stopping criterion.

After the adaptive procedure has terminated and the model configuration has been fixed, the final surrogate is evaluated on the independent test subset.
The effectiveness of the uncertainty-guided adaptive strategy is evaluated in Section~\ref{sec:verification} through comparisons with random sample selection and analyses of surrogate convergence.

The adaptive procedure demonstrated here is restricted to pool-based sample selection within the existing simulation database; no new \textsc{Varyped} equilibria, \textsc{Hypnotoad} meshes, or \textsc{BOUT++} simulations are generated automatically. Nevertheless, the same probabilistic framework provides the foundation for future closed-loop simulation campaigns in which surrogate uncertainty is used to trigger additional high-fidelity simulations and iteratively expand the database.

\section{Simulation Database Characteristics}
\label{sec:statistics}


\begin{figure*}[t]
    \centering
    \includegraphics[width=\textwidth]{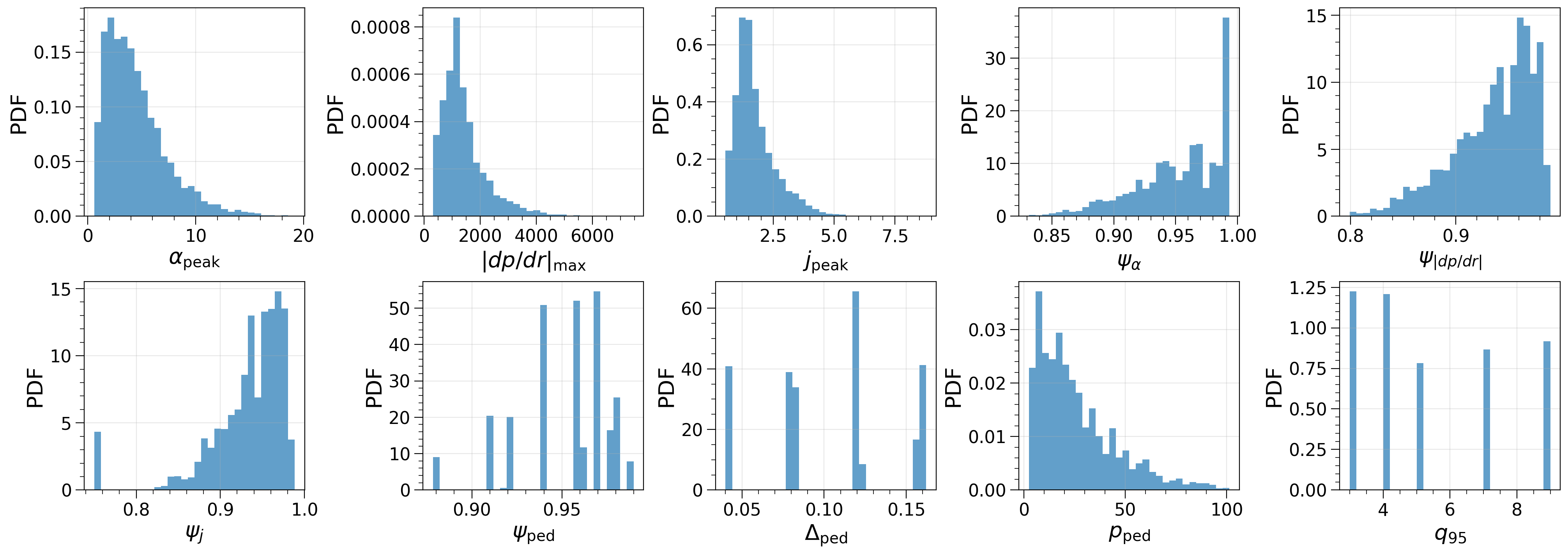}
    \caption{\textbf{Probability density functions (PDFs) of the ten input features used for Gaussian Process training.} The distributions characterize the realized parameter-space coverage of the quality-controlled single-shape DIII-D equilibrium database. The features include the peak normalized pressure-gradient parameter $\alpha_{\rm peak}$, the maximum pressure gradient $|dp/dr|_{\max}$, the peak current density $j_{\rm peak}$, the radial locations of these quantities ($\psi_\alpha$, $\psi_{|dp/dr|}$, and $\psi_j$), the pedestal location $\psi_{\rm ped}$, the pedestal width $\Delta_{\rm ped}$, the pedestal height $p_{\rm ped}$, and the edge safety factor $q_{95}$. Together, these distributions illustrate the range of pedestal characteristics represented in the simulation database used for surrogate-model development.}
    \label{fig:database_distribution}
\end{figure*}

\begin{figure*}[t]
    \centering
    \includegraphics[width=0.65\textwidth]{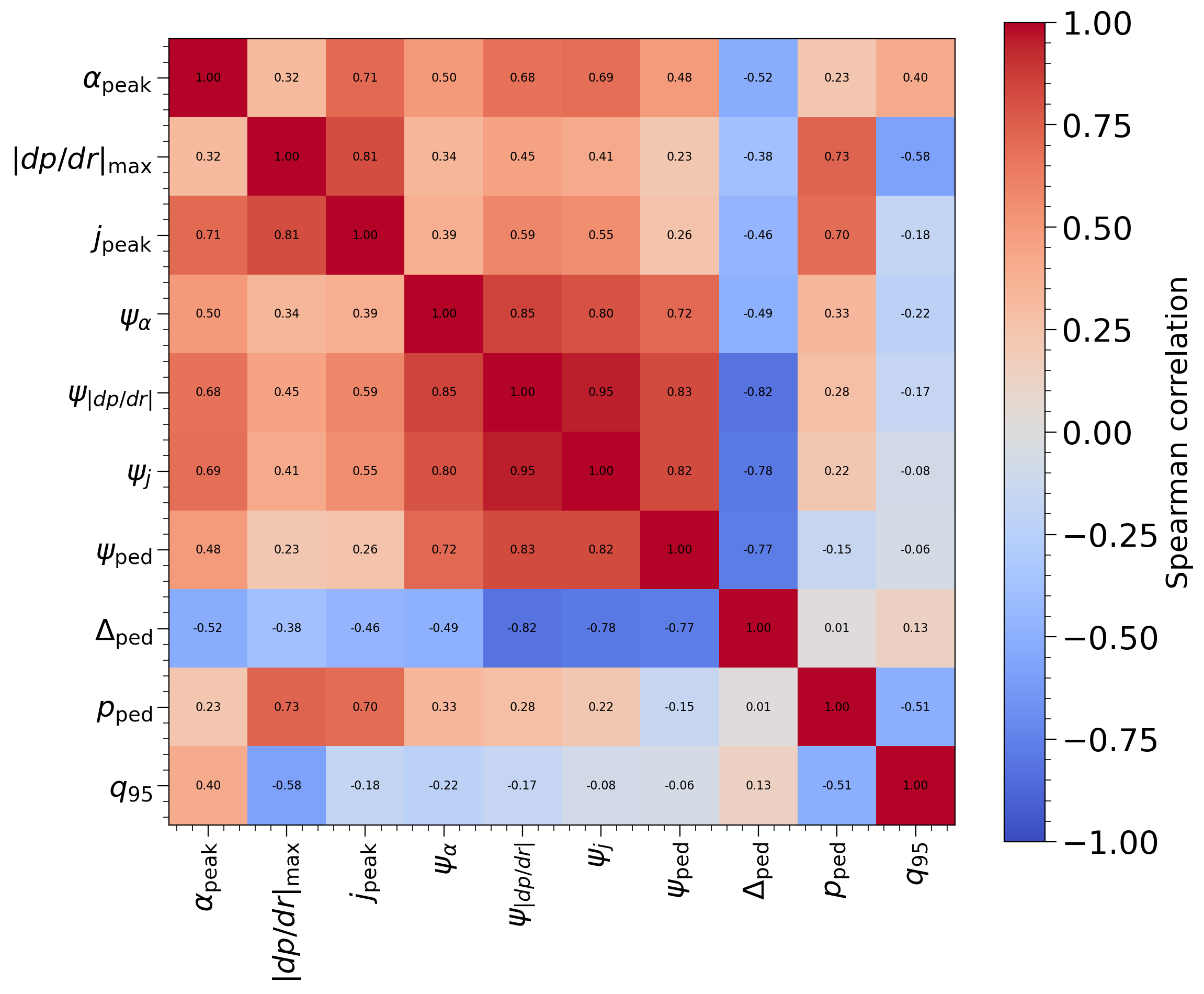}
    \caption{\textbf{Spearman rank correlation matrix for the ten Gaussian Process input features.} The heatmap illustrates the pairwise monotonic relationships among the equilibrium-derived input features used for surrogate training. The peak normalized pressure-gradient parameter $\alpha_{\rm peak}$, maximum pressure gradient $|dp/dr|_{\max}$, and peak current density $j_{\rm peak}$ exhibit moderate to strong positive correlations. The radial locations of the pressure-gradient peak, current-density peak, and pedestal ($\psi_\alpha$, $\psi_{|dp/dr|}$, $\psi_j$, and $\psi_{\rm ped}$) are highly correlated with one another, whereas the pedestal width $\Delta_{\rm ped}$ is strongly anti-correlated with these location parameters. In contrast, the pedestal height $p_{\rm ped}$ and edge safety factor $q_{95}$ exhibit relatively weak correlations with most of the remaining features. These relationships reflect the underlying Varyped parameterization and identify potentially redundant input features for the recursive feature-selection procedure.
    }
    \label{fig:spearman}
\end{figure*}

This section summarizes the characteristics of the quality-controlled simulation database used for Gaussian Process surrogate development. Following equilibrium reconstruction, mesh generation, \textsc{BOUT++} simulations, and quality-control filtering, the final database contains 3,869 equilibria together with their corresponding ideal-MHD and ideal-plus-diamagnetic linear-stability calculations.

As described in Section~\ref{database}, the quality-controlled database was constructed from the original 7,992 requested Varyped equilibrium configurations through successive equilibrium reconstruction, mesh generation, \textsc{BOUT++} simulations, and quality-control filtering. Each retained equilibrium contributes two sixteen-mode growth-rate spectra corresponding to the ideal-MHD and ideal-plus-diamagnetic physics models, yielding

\[
3,\!869\times32 = 123,\!808
\]
mode-resolved \textsc{BOUT++} linear-stability calculations.

\subsection{Equilibrium-Parameter Coverage and Correlations}

Figure~\ref{fig:database_distribution} summarizes the probability density functions of the ten equilibrium-derived features extracted from the simulation database. These include the peak normalized pressure-gradient parameter $\alpha_{\rm peak}$, maximum pressure gradient $|dp/dr|_{\max}$, peak current density $j_{\rm peak}$, the radial locations of these quantities ($\psi_{\alpha}$, $\psi_{|dp/dr|}$, and $\psi_j$), pedestal location $\psi_{\rm ped}$, pedestal width $\Delta_{\rm ped}$, pedestal height $p_{\rm ped}$, and the edge safety factor $q_{95}$.

The resulting feature distributions characterize the realized parameter-space coverage following equilibrium-convergence and quality-control filtering. They span the populated regions of the pedestal parameter space sampled by the present single-shape Varyped scan and define the input domain used for surrogate-model development.

The relationships among the input features are summarized by the Spearman rank-correlation matrix shown in Figure~\ref{fig:spearman}. Several expected physical correlations are observed. The radial locations of the pressure-gradient peak, current-density peak, and pedestal center ($\psi_{\alpha}$, $\psi_{|dp/dr|}$, $\psi_j$, and $\psi_{\rm ped}$) exhibit strong positive correlations, reflecting the close coupling between pressure and current profiles in the pedestal region. Likewise, the maximum pressure gradient and peak current density are positively correlated, whereas the pedestal width is strongly anti-correlated with these location parameters.

These relationships are consistent with the underlying \textsc{Varyped} parameterization and identify potentially redundant input features that are subsequently examined through the recursive feature-selection procedure described in Section~\ref{sec:feature}.

\subsection{Linear Growth-Rate Characteristics}

\begin{figure*}[t]
    \centering

    \begin{overpic}[width=0.48\textwidth]{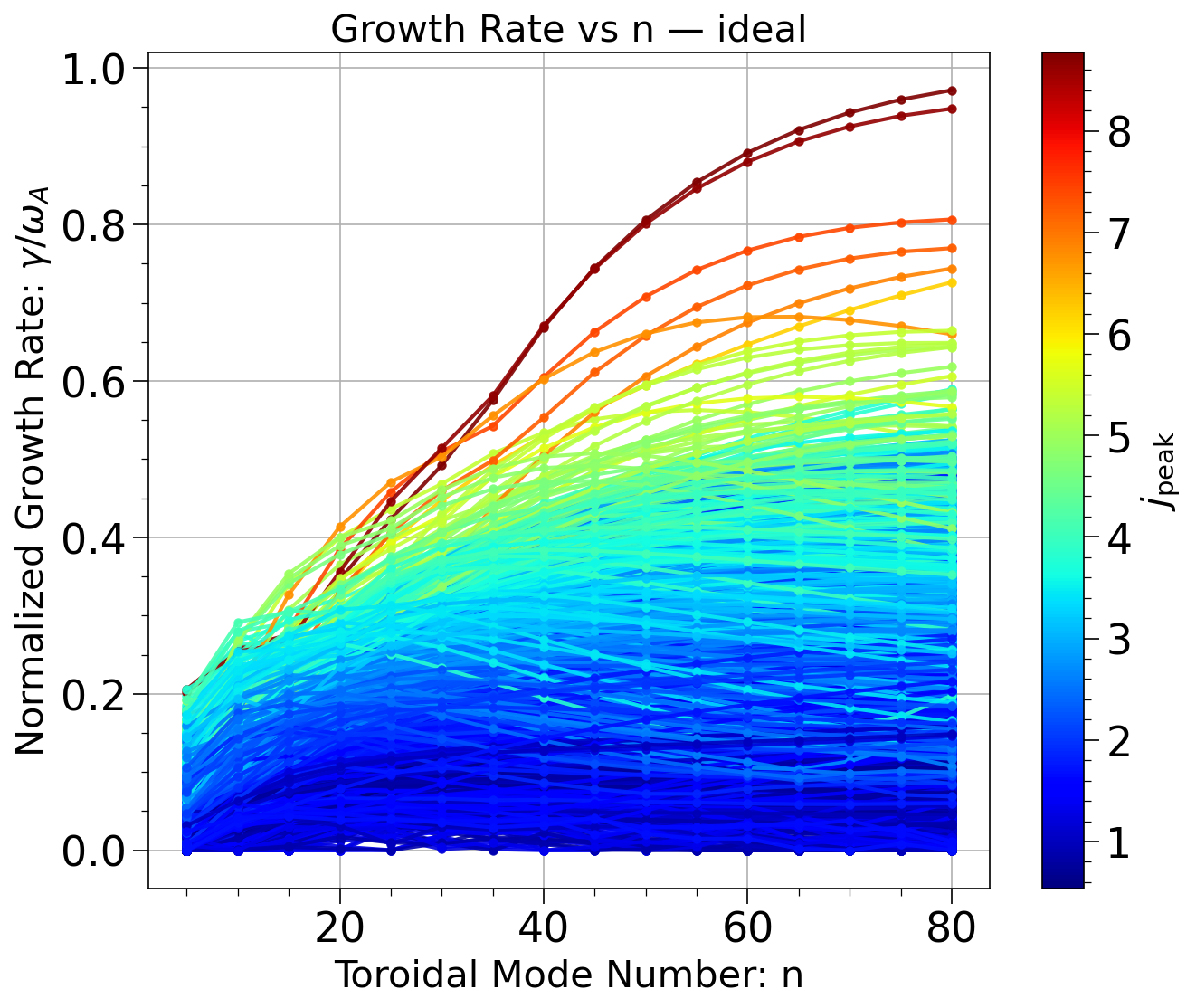}
        \put(15,72){\large\textbf{(a)}}
    \end{overpic}
    \hfill
    \begin{overpic}[width=0.48\textwidth]{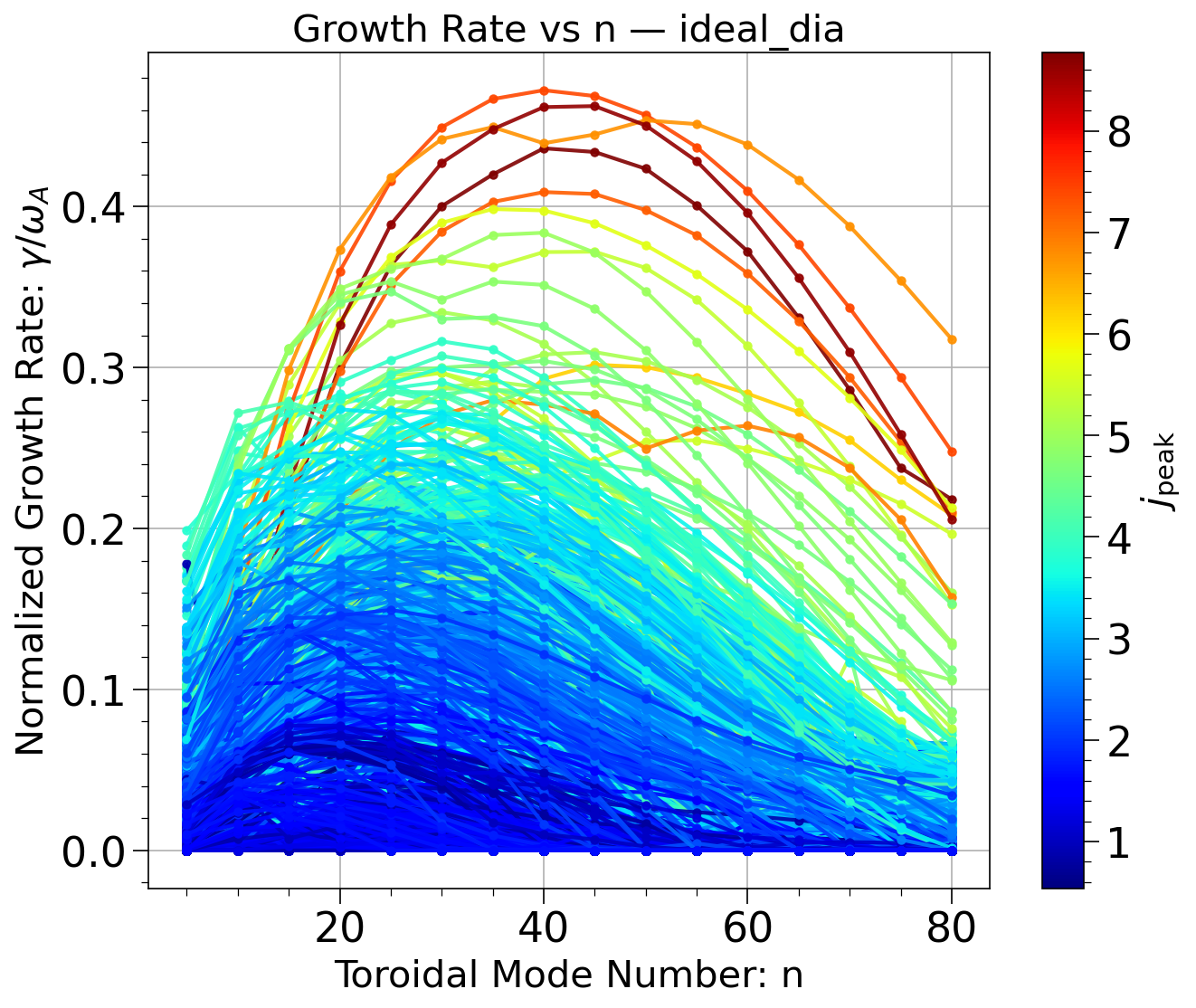}
        \put(15,72){\large\textbf{(b)}}
    \end{overpic}

    \vspace{0.3cm}

    \begin{overpic}[width=0.48\textwidth]{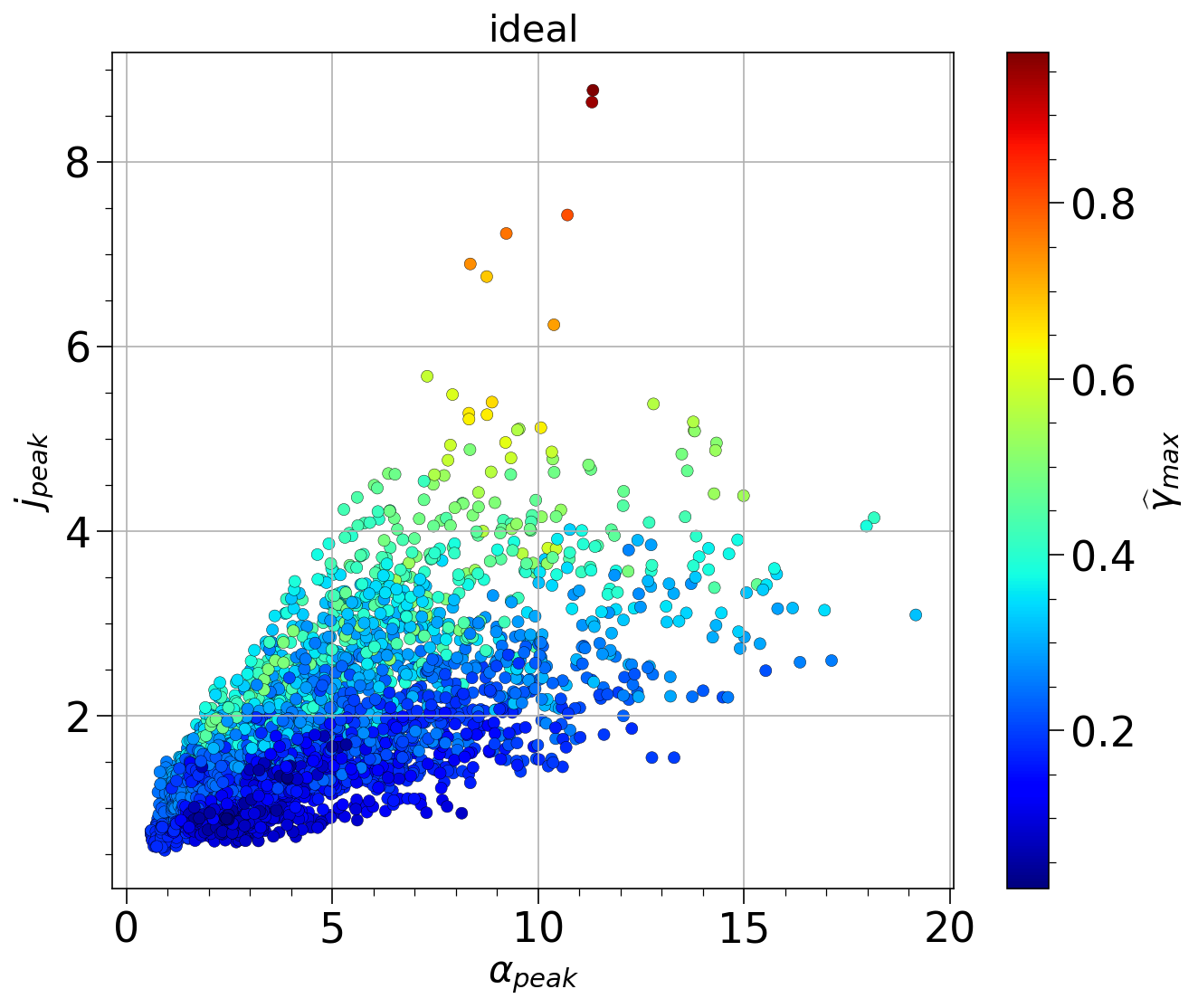}
        \put(15,72){\large\textbf{(c)}}
    \end{overpic}
    \hfill
    \begin{overpic}[width=0.48\textwidth]{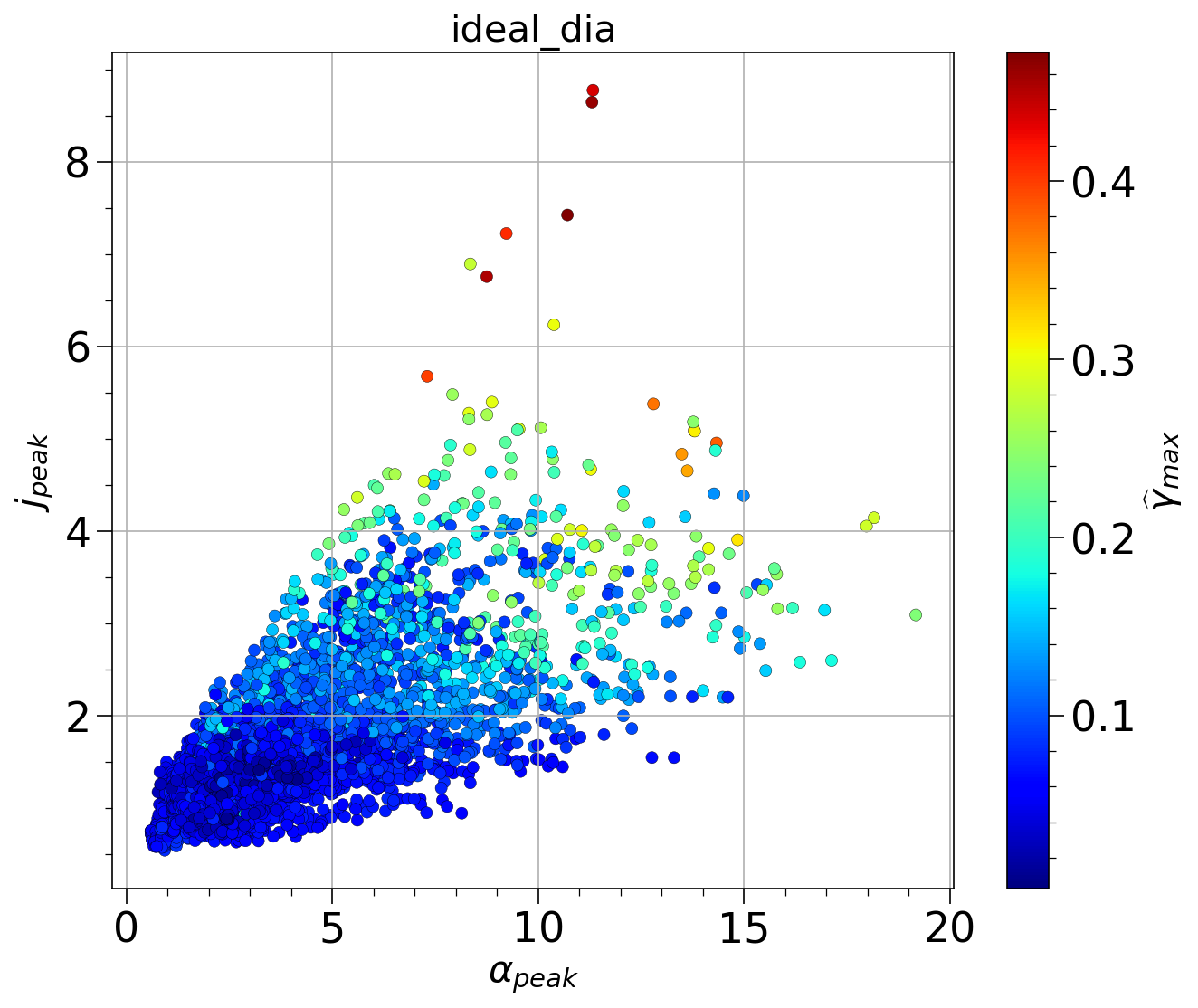}
        \put(15,72){\large\textbf{(d)}}
    \end{overpic}

    \caption{\textbf{Linear-stability characteristics of the ELMO simulation database.} Panels \textbf{(a)} and \textbf{(b)} show the linear growth-rate spectra from the ideal-MHD and ideal-plus-diamagnetic \textsc{BOUT++} simulations, respectively. Each curve corresponds to a single equilibrium and is colored by the peak current density, $j_{\rm peak}$. In the ideal-MHD case, the most unstable mode changes non-monotonically with increasing $j_{\rm peak}$: high-$n$ modes dominate at low current density, the dominant mode shifts to intermediate $n$ at moderate current density, and high-$n$ modes become dominant again at larger current density. In contrast, the inclusion of diamagnetic effects suppresses the high-$n$ modes and shifts the dominant instability toward intermediate toroidal mode numbers over much of the parameter space. Panels \textbf{(c)} and \textbf{(d)} show the joint distribution of $\alpha_{\rm peak}$ and $j_{\rm peak}$, with marker color indicating the maximum linear growth rate, $\widehat\gamma_{\max}$. The distributions illustrate the progression from weakly unstable to strongly unstable pedestal conditions as the pressure gradient and current density increase.
    Together, these results demonstrate the wide range of pedestal pressure-gradient, current-density, and linear-stability characteristics represented within the quality-controlled simulation database used for Gaussian Process surrogate development.}

    \label{fig:growth_database}
\end{figure*}

Figure~\ref{fig:growth_database} summarizes the linear-stability characteristics of the simulation database. Panels~(a) and~(b) present the growth-rate spectra obtained from the ideal-MHD and ideal-plus-diamagnetic \textsc{BOUT++} calculations, respectively. The database spans a broad range of stability behavior, ranging from nearly stable equilibria to strongly unstable cases characterized by rapidly growing high-$n$ modes.

The inclusion of diamagnetic effects reduces the overall growth rates and shifts the dominant instability toward intermediate toroidal mode numbers. Panels~(c) and~(d) illustrate the relationship between the peak normalized pressure-gradient parameter ($\alpha_{\rm peak}$), peak current density ($j_{\rm peak}$), and the corresponding maximum linear growth rate ($\widehat\gamma_{\max}$). Together, these distributions demonstrate the broad range of equilibrium and stability characteristics represented within the quality-controlled simulation database used for Gaussian Process surrogate development.


\section{Surrogate Verification and Independent Testing}
\label{sec:verification}

Feature selection and adaptive-training convergence are evaluated using the fixed validation set described in Section~\ref{sec:train}. These validation cases are excluded from Gaussian Process fitting and adaptive sample acquisition but are intentionally used during model development for feature-selection decisions and determination of the adaptive stopping point.

Final prediction performance is assessed separately using the independent test subset associated with each of the five random data-partition realizations. Within each realization, the test equilibria are excluded from Gaussian Process fitting, feature selection, adaptive sample acquisition, learning-curve construction, and stopping decisions. Test metrics are not used in any model-development decision, and only results from the frozen final surrogate are reported as independent-test performance. Unless otherwise stated, final test metrics are calculated separately for each realization and summarized by their mean and sample standard deviation over the five realizations. Accordingly, the final $R^2$, RMSE, MAE, and MRE metrics, together with the results for $\widehat{\gamma}_{\max}$ and $n_{\max}$, provide an independent estimate of surrogate generalization performance.

Throughout this section, surrogate predictions are compared directly with the corresponding \textsc{BOUT++} linear-stability calculations to assess prediction accuracy, latent posterior uncertainty and its calibration, and computational efficiency. Results associated with feature selection and convergence are reported on the validation set, whereas the final prediction and error statistics are reported on the independent test subsets.

This section first presents the results of the physics-informed feature-selection procedure, followed by the convergence behavior of the adaptive-training strategy. The surrogate is then evaluated through comparisons of the predicted growth-rate spectra, maximum linear growth rates, dominant unstable toroidal modes, prediction errors, latent posterior-uncertainty calibration, and computational performance.

\subsection{Feature-Selection Results}


\begin{figure*}[t]
  \centering
  \includegraphics[width=0.9\textwidth]{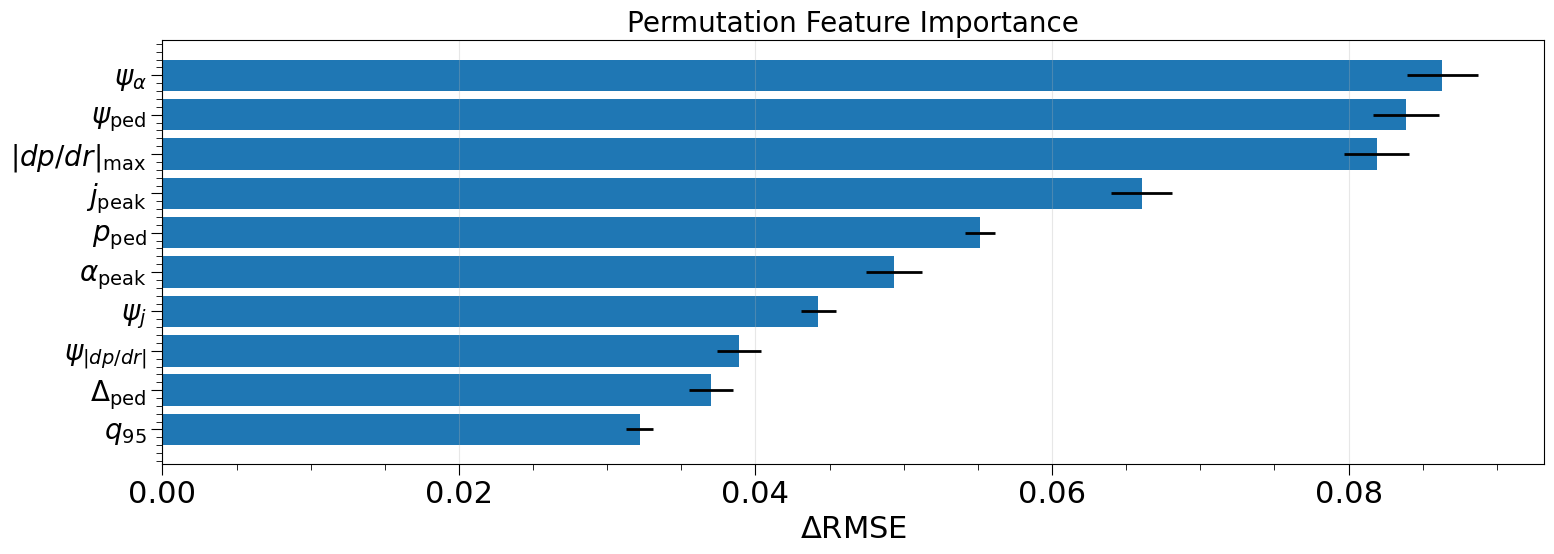}
  \caption{\textbf{Permutation feature importance of the original ten-feature Gaussian Process surrogate.} Feature importance is quantified by the increase in the prediction root-mean-square error ($\Delta$RMSE) after randomly permuting each input feature while keeping all remaining features fixed. Error bars denote the standard deviation over repeated permutations. The radial location of the peak normalized pressure gradient ($\psi_{\alpha}$), the maximum pressure gradient ($|dp/dr|_{\max}$), and the pedestal location ($\psi_{\rm ped}$) produce the largest increases in prediction error, indicating that these features have the greatest individual influence on the original ten-feature surrogate. In contrast, the edge safety factor ($q_{95}$) exhibits the smallest increase in prediction error and is therefore identified as the first candidate for recursive feature elimination. The final feature-selection decisions are based on the combined evaluation of permutation importance, feature correlation, multicollinearity, and surrogate retraining, as described in Section~\ref{sec:feature}.}
  \label{fig:feature_importance}
\end{figure*}

To construct an efficient and physically interpretable surrogate model, the importance of each candidate input feature was first evaluated using permutation feature importance applied to the original ten-feature Gaussian Process surrogate. For each feature, its values were randomly permuted over the fixed held-out validation set while all remaining input features were held fixed. The resulting increase in prediction error, measured by the increase in the root-mean-square error (RMSE), was used as the feature-importance metric.

Figure~\ref{fig:feature_importance} summarizes the permutation feature importance obtained for the original ten-feature surrogate. The radial location of the peak normalized pressure-gradient parameter ($\psi_{\alpha}$), the maximum pressure gradient ($(dp/dr)_{\rm max}$), and the pedestal-center location ($\psi_{\rm ped}$) produce the largest increases in prediction error when permuted, indicating that these features have the greatest individual influence on the predictive performance of the original surrogate. In contrast, the edge safety factor ($q_{95}$) produces the smallest increase in prediction error, identifying it as the first candidate for recursive feature elimination.

Feature reduction was subsequently performed using the physics-informed recursive backward feature-elimination procedure described in Section~\ref{sec:feature}. Following removal of $q_{95}$, the Gaussian Process surrogate was retrained using the remaining nine features. Permutation feature importance, Spearman rank correlation, and variance inflation factors were then recomputed for the retrained surrogate to identify the next candidate features for removal. This second evaluation identified the peak normalized pressure-gradient parameter ($\alpha_{\rm peak}$) and the pedestal-center location ($\psi_{\rm ped}$) as candidate removable features.

The two candidates were subsequently evaluated independently by constructing separate eight-feature surrogate models, one excluding $\alpha_{\rm peak}$ and the other excluding $\psi_{\rm ped}$. For each candidate, the complete adaptive-training workflow was repeated over five random data-partition realizations, and surrogate performance was evaluated exclusively on the corresponding validation subsets. Removing $\alpha_{\rm peak}$ produced a noticeable reduction in the mean coefficient of determination, demonstrating that this feature contains unique predictive information essential for accurate stability prediction. In contrast, removing $\psi_{\rm ped}$ resulted in essentially no degradation in predictive performance, indicating that its contribution is largely redundant with the remaining pedestal-location features. Consequently, $\psi_{\rm ped}$ was removed from the final surrogate model, whereas $\alpha_{\rm peak}$ was retained.

The recursive feature-elimination procedure terminated after the resulting eight-feature model satisfied the prescribed acceptance criterion. As shown in the following subsection, the reduced-feature surrogate preserves essentially the same predictive accuracy as the original ten-feature model while reducing training cost and improving model interpretability.

\subsection{Learning-Curve and Training-Set Convergence}

\begin{figure*}[t]
  \centering

  \begin{overpic}[width=0.48\textwidth]{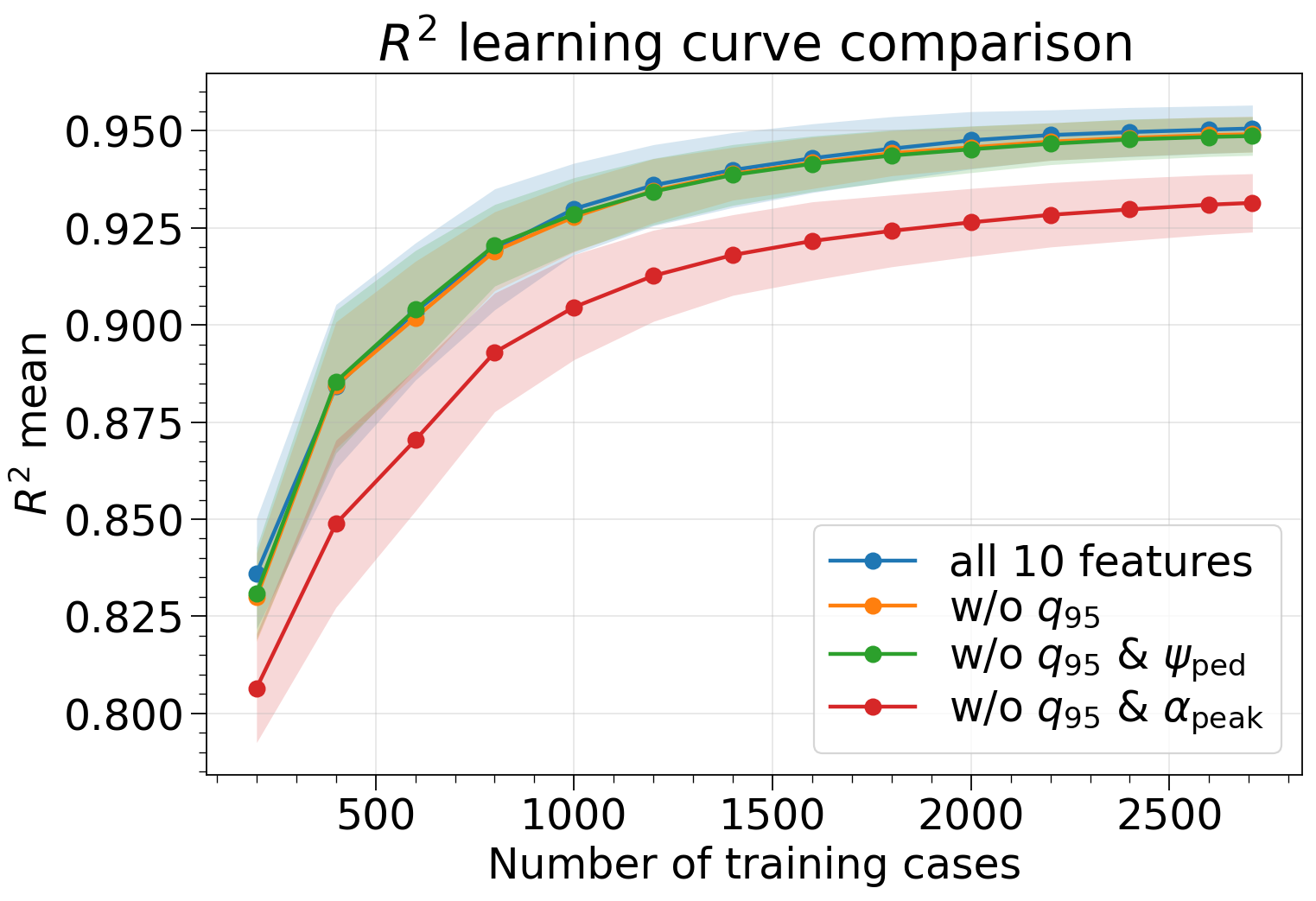}
    \put(85,42){\large\textbf{(a)}}
  \end{overpic}
  \hfill
  \begin{overpic}[width=0.48\textwidth]{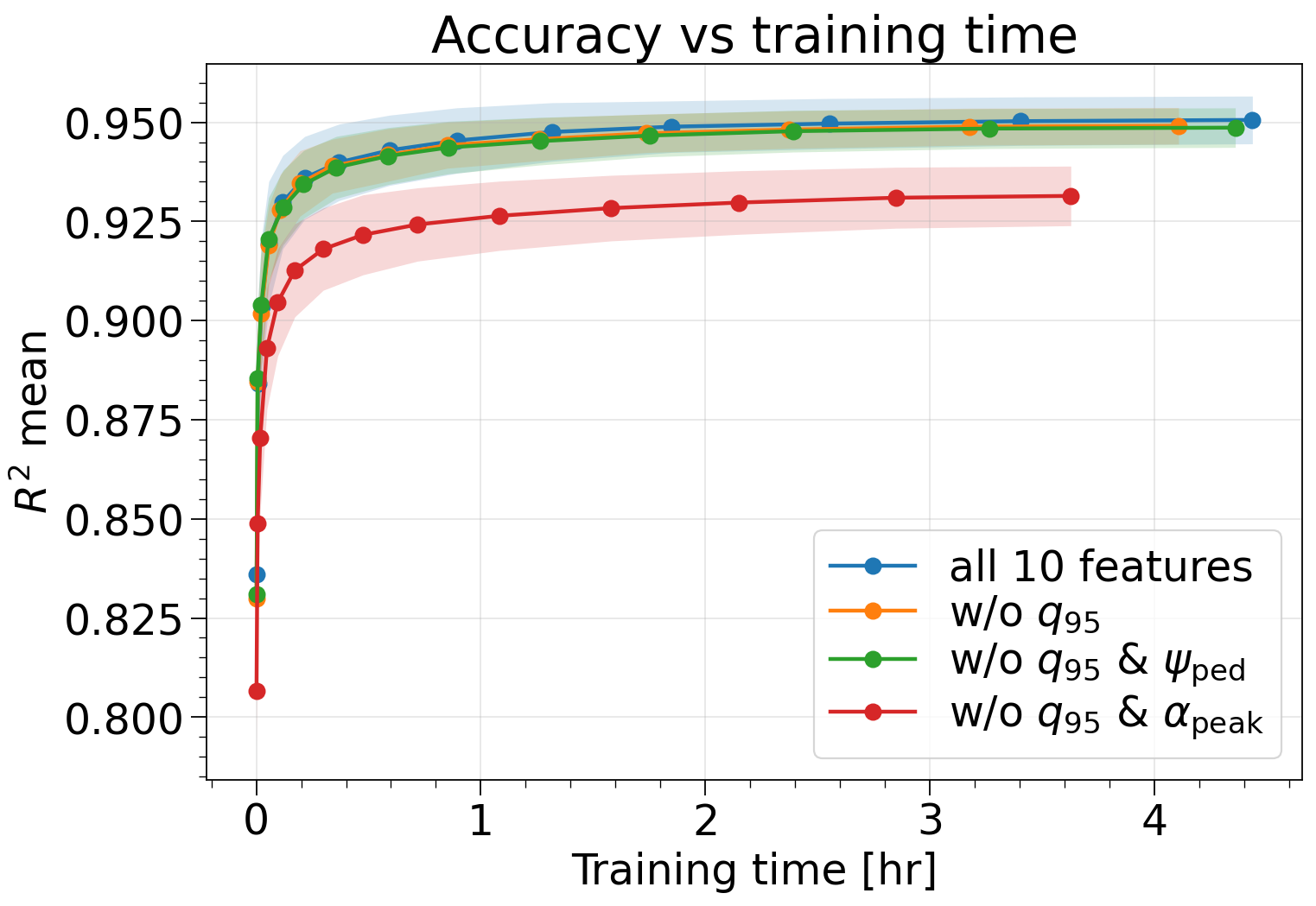}
    \put(85,42){\large\textbf{(b)}}
  \end{overpic}

  \caption{\textbf{Learning curves and feature-selection performance of the Gaussian Process surrogate.}
  \textbf{(a)} Mean validation $R^2$ versus the number of training cases for different feature sets. Starting from the original ten-feature model, removing the edge safety factor ($q_{95}$) has a negligible impact on validation accuracy while modestly reducing computational cost. Subsequent removal of $\psi_{\rm ped}$ yields an eight-feature surrogate that preserves essentially the same validation performance as the original model, whereas removing $\alpha_{\rm peak}$ instead results in a noticeable degradation in model accuracy.
  \textbf{(b)} Mean validation $R^2$ versus cumulative training time. Solid curves denote the mean validation performance over five random data-partition realizations, and shaded regions denote one sample standard deviation.
  All models exhibit rapid improvement with increasing training data before approaching convergence near $R^2\approx0.95$ after approximately 2,000 training samples. The results demonstrate that the final eight-feature surrogate preserves the validation performance of the original model while improving computational efficiency. The reported learning curves are evaluated on the fixed validation subset of each realization.
}
  \label{fig:learning_curve}
\end{figure*}

The convergence behavior of the Gaussian Process surrogate is evaluated using the pool-based adaptive sample-selection procedure described in Section~\ref{sec:UQ}. Figure~\ref{fig:learning_curve} summarizes the evolution of the mean surrogate accuracy as the training set is progressively expanded. Reported learning curves represent the mean validation performance over the fixed validation subsets of the five random data-partition realizations, with the shaded regions indicating one sample standard deviation.

Figure~\ref{fig:learning_curve}(a) compares the validation accuracy obtained using the original ten-feature model and the reduced feature sets generated by the physics-informed recursive feature-elimination procedure. Beginning with approximately 200 randomly selected training samples, the validation accuracy improves rapidly during the early stages of training, reaching approximately $R^2\approx0.94$ with roughly 1,600 training samples. Beyond approximately 2,000 training samples, the learning curves begin to plateau, approaching a converged accuracy of $R^2\approx0.95$. Additional samples drawn from the existing simulation database provide only modest improvements, indicating diminishing returns from further expansion of the current training set.

For the retrospective learning-curve and adaptive-versus-random comparisons in Figs.~\ref{fig:learning_curve} and~\ref{fig:adaptive_vs_random}, training was continued beyond the operational stopping point toward the full adaptive-training pool. These extended curves are presented only to characterize convergence and compare the two sampling strategies over a common range of training-set sizes. Results obtained after the operational stopping point were not used to revise the feature set, stopping criterion, or other model-development choices.

The influence of recursive feature elimination is also evident. Removing the edge safety factor ($q_{95}$) produces almost no change in validation accuracy while reducing the computational cost of surrogate training. Following retraining of the nine-feature model, subsequent removal of the pedestal-center location ($\psi_{\rm ped}$) likewise preserves essentially the same validation performance as the original model. In contrast, removing the peak normalized pressure-gradient parameter ($\alpha_{\rm peak}$) produces a noticeable degradation in surrogate performance, demonstrating that this feature provides unique predictive information that cannot be recovered from the remaining inputs.

Figure~\ref{fig:learning_curve}(b) presents the corresponding validation accuracy as a function of cumulative training time. The final eight-feature surrogate consistently achieves essentially the same validation accuracy as the original ten-feature model while requiring less computational effort. It therefore provides an effective compromise among predictive accuracy, computational efficiency, and physical interpretability.

The convergence behavior observed in Fig.~\ref{fig:learning_curve} indicates that additional training samples from the existing database provide progressively smaller improvements in surrogate performance. This saturation suggests that further gains are likely to require more informative training samples or expansion of the underlying equilibrium database rather than simply increasing the size of the current training set.

The learning curves in Fig.~\ref{fig:learning_curve} represent validation performance used during model development and should therefore not be interpreted as independent-test results. Final surrogate accuracy is evaluated separately on the independent 15\% test subsets in Sections~\ref{sec:growthrate} and~\ref{sec:errors}.

\subsection{Adaptive Training Performance}

\begin{figure*}[t]
  \centering
  \includegraphics[width=0.68\textwidth]{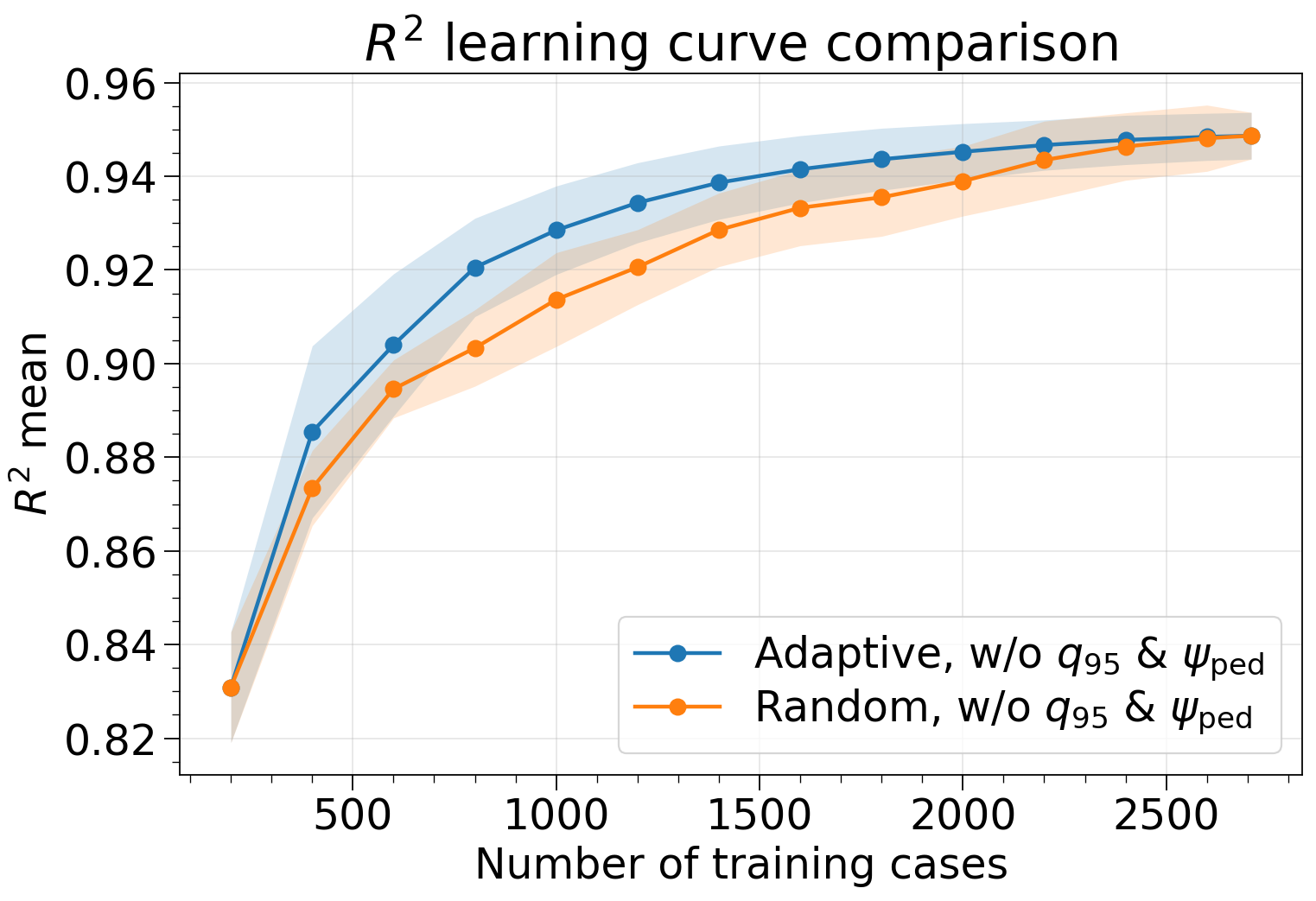}
  \caption{\textbf{Comparison between uncertainty-guided adaptive training and random training-set selection.}
  Both strategies begin from the same randomly selected initial training subset. As the training set is expanded, the adaptive strategy selects new samples from the candidate pool according to the mean Gaussian Process latent posterior standard deviation across all surrogate output quantities, whereas the random strategy selects additional samples randomly. The acquisition scores are evaluated in standardized output space. Solid curves show the mean validation $R^2$ over five random data-partition realizations, and shaded regions denote one sample standard deviation. The adaptive strategy generally achieves higher mean validation accuracy at intermediate training-set sizes, indicating improved sample efficiency. As the training set approaches the full adaptive-training pool, the two strategies converge because they ultimately incorporate essentially the same available cases.}
  \label{fig:adaptive_vs_random}
\end{figure*}

One of the principal advantages of Gaussian Process Regression is that it provides model-based posterior uncertainty together with the surrogate prediction. In the present study, this uncertainty is used as a relative acquisition score to guide pool-based adaptive sample selection within the existing simulation database, with additional training samples selected according to the mean latent posterior standard deviation across the 32 output channels.

Figure~\ref{fig:adaptive_vs_random} compares uncertainty-guided sample selection with random training-set selection over five random data-partition realizations. At each training-set size, the solid curves represent the mean validation $R^2$, while the shaded regions indicate one sample standard deviation. Both strategies begin from the same randomly selected initial training subset and therefore exhibit essentially identical validation accuracy at the initial training size. As the training set is expanded, uncertainty-guided sampling preferentially selects cases with larger Gaussian Process posterior uncertainty, whereas the random strategy adds cases without regard to the surrogate uncertainty estimate. Over this intermediate training regime, the adaptive strategy generally achieves a higher mean validation $R^2$ for a given number of training cases, indicating improved sample efficiency. As the training-set size approaches the full adaptive-training pool, the two strategies converge because both ultimately incorporate essentially the same available cases. Consequently, the benefit of uncertainty-guided sampling is most evident at intermediate training-set sizes rather than at either endpoint.

Although the adaptive strategy generally achieves higher mean validation accuracy over the intermediate training regime, the improvement is modest relative to the realization-to-realization variability, as indicated by the overlap of the variability bands. This behavior is consistent with the structure of the realized simulation database. Although equilibrium-convergence filtering introduces structured gaps into the sampled parameter space, the populated regions remain sufficiently redundant that uncertainty-guided sample selection provides only limited improvement over random selection within the existing candidate pool. Accordingly, the present results support uncertainty-guided selection as a useful strategy for improving sample efficiency without demonstrating consistent superiority over random selection across the full training range.

The learning curves also reveal diminishing returns as the training set increases. Beyond approximately 2,000 training samples, the validation accuracy begins to saturate, with additional training samples producing only marginal improvements in $R^2$ while substantially increasing the surrogate training time. This behavior suggests that simply increasing the size of the current training set is unlikely to substantially improve surrogate performance. Instead, further gains are expected to require more informative samples or expansion of the simulation database into comparatively under-represented regions of parameter space through additional high-fidelity simulations.

As the ELMO database is extended to include multiple plasma shapes, broader pedestal parameter ranges, and additional physics models, including resistive and drift-Alfv\'en effects, uncertainty-guided sample selection may become increasingly valuable. In particular, model-based posterior uncertainty can provide a relative criterion for identifying candidate regions in which additional high-fidelity simulations may improve surrogate coverage. Because the present posterior uncertainties are not fully calibrated, their use in closed-loop database expansion should be accompanied by calibration monitoring, out-of-distribution diagnostics, or comparisons with alternative acquisition criteria.

It should be emphasized that the present study demonstrates pool-based adaptive sample selection using an existing simulation database rather than a fully automated adaptive simulation workflow. No new \textsc{Varyped} equilibria, \textsc{Hypnotoad} meshes, or \textsc{BOUT++} calculations are generated during the adaptive procedure. Extension to a closed-loop simulation-to-surrogate workflow, in which a validated acquisition criterion is used to trigger additional high-fidelity simulations, remains a subject for future work.

\subsection{Growth-Rate Prediction}
\label{sec:growthrate}

\begin{figure*}[t]
  \centering
  \begin{overpic}[width=0.95\textwidth]{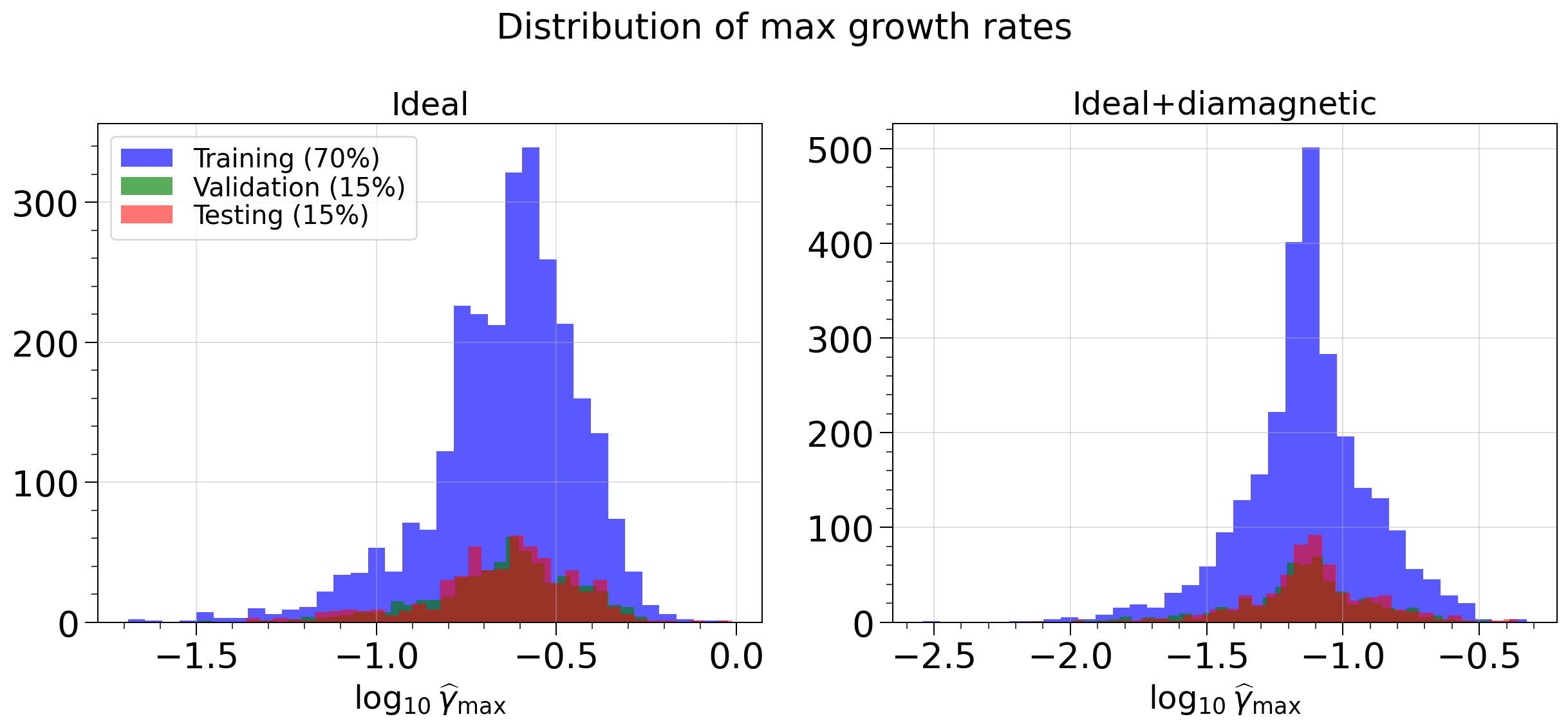}
    \put(41,35){\large\textbf{(a)}}
    \put(60,35){\large\textbf{(b)}}
  \end{overpic}
  \caption{\textbf{Distributions of the maximum linear growth rates in the adaptive-training pool, validation, and independent test subsets for the ideal-MHD (a) and ideal-plus-diamagnetic (b) models.} Histograms are shown for the adaptive-training pool (70\%, blue), validation set (15\%, green), and independent test set (15\%, red) using $\log_{10}\widehat\gamma_{\max}$. The three subsets exhibit similar distributions for both physics models, indicating that the validation and test subsets provide representative coverage of the stability behavior contained in the quality-controlled simulation database.}
  \label{fig:growth_rate_pdf}
\end{figure*}

\begin{figure*}[t]
  \centering
  \begin{overpic}[width=0.9\textwidth]{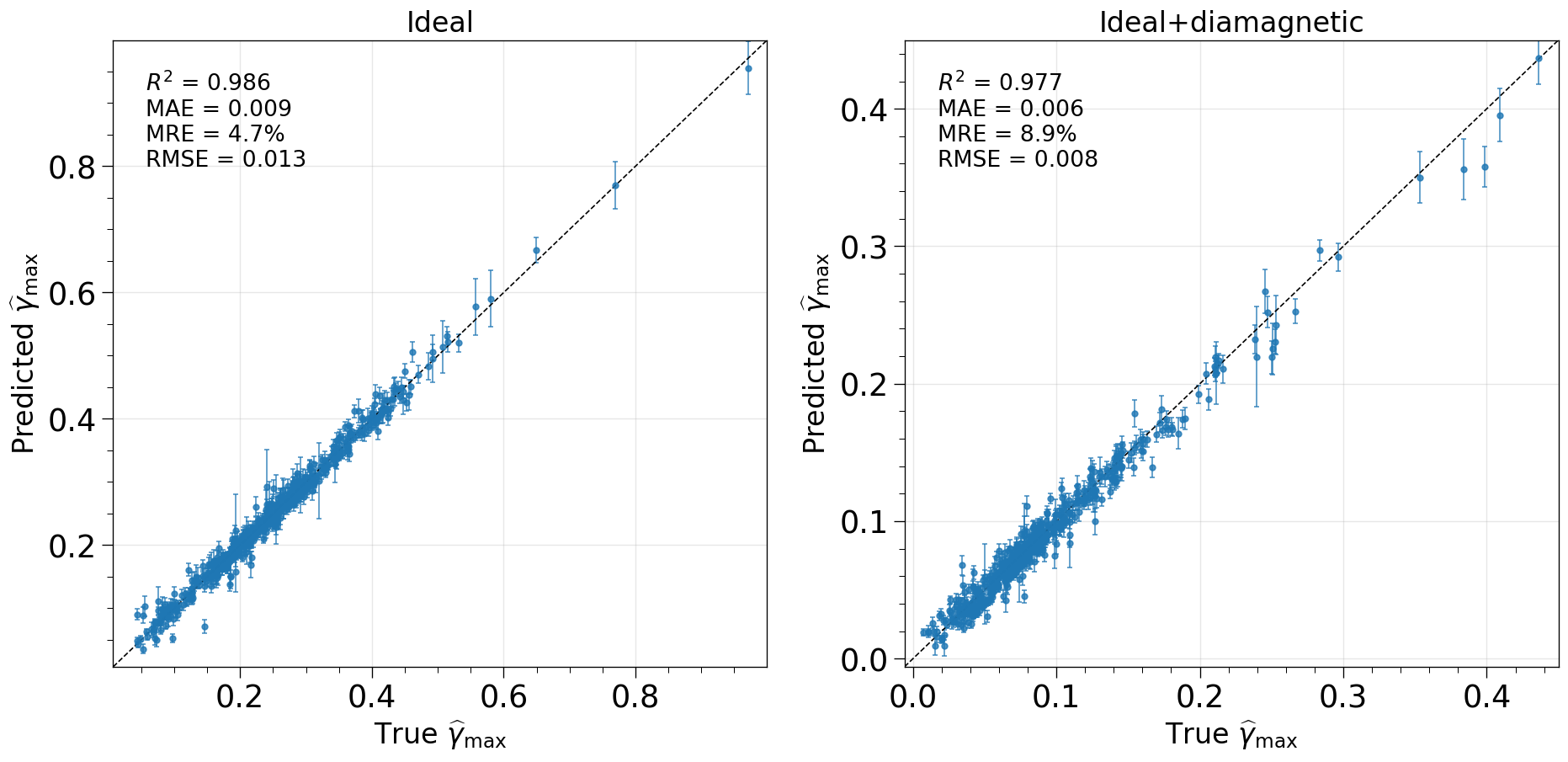}
    \put(43,8){\large\textbf{(a)}}
    \put(93,8){\large\textbf{(b)}}
  \end{overpic}
  \caption{\textbf{Parity plots comparing Gaussian Process predictions and \textsc{BOUT++}-computed maximum linear growth rates for the ideal-MHD \textbf{(a)} and ideal-plus-diamagnetic \textbf{(b)} models.}
  The displayed predictions correspond to the independent test subset from the designated realization with random seed 456. The dashed line denotes perfect agreement between the predicted and simulated values of $\widehat{\gamma}_{\max}$. The error bar for each equilibrium represents one latent posterior standard deviation for the output channel corresponding to the surrogate-predicted peak mode; it is not a separately propagated uncertainty for the maximum operation. Final $\widehat{\gamma}_{\max}$ metrics were calculated separately on the independent test subset of each realization and then summarized across the five realizations. The ideal-MHD model achieved $R^2=0.978\pm0.013$, MAE $=0.00938\pm0.00059$, RMSE $=0.0153\pm0.0046$, and MRE $=(4.90\pm0.24)\%$. The ideal-plus-diamagnetic model achieved $R^2=0.966\pm0.009$, MAE $=0.00588\pm0.00020$, RMSE $=0.00887\pm0.00092$, and MRE $=(9.17\pm0.62)\%$. The reported $\pm$ values denote one sample standard deviation across the five random data-partition realizations.}
  \label{fig:gp_results}
\end{figure*}

\begin{figure*}[t]
  \centering
  \begin{overpic}[width=0.95\textwidth]{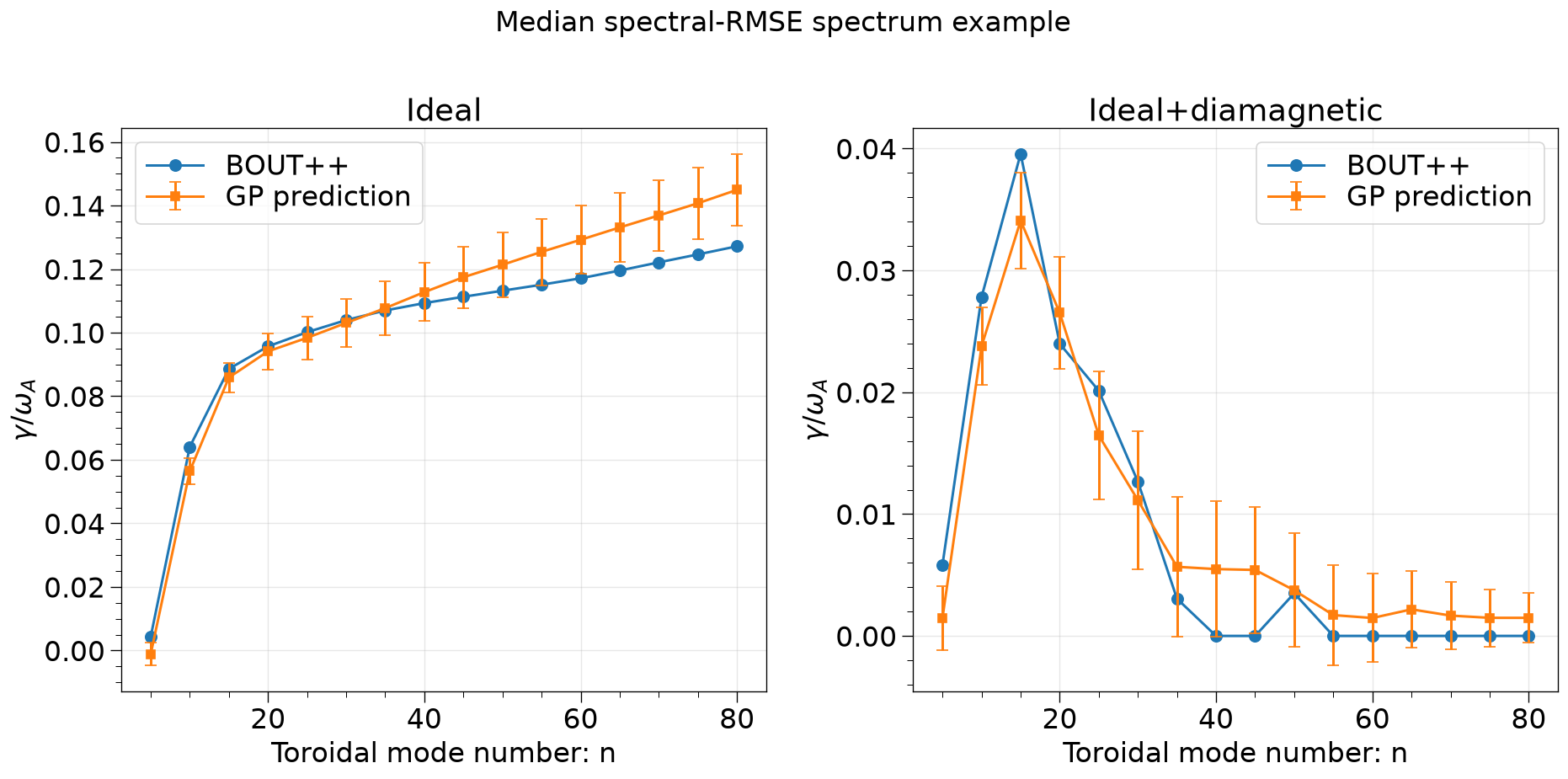}
    \put(42,8){\large\textbf{(a)}}
    \put(60,8){\large\textbf{(b)}}
  \end{overpic}
  \caption{\textbf{Gaussian Process prediction of the complete growth-rate spectrum for a median-error independent-test equilibrium.}
  Blue circles denote the normalized growth rates computed by \textsc{BOUT++}, and orange squares denote the Gaussian Process posterior-mean predictions. Error bars indicate one Gaussian Process latent posterior standard deviation. The equilibrium was selected objectively from the seed-456 independent test subset using the combined spectral-RMSE criterion defined in Eqs.~\eqref{eq:model_spectral_rmse} and~\eqref{eq:combined_spectral_rmse}. Panels \textbf{(a)} and \textbf{(b)} show the ideal-MHD and ideal-plus-diamagnetic spectra, respectively.}
  \label{fig:growth_spectrum_example}
\end{figure*}

Final surrogate accuracy was evaluated exclusively on the independent test subsets. Figure~\ref{fig:growth_rate_pdf} compares the distributions of $\widehat{\gamma}_{\max}$ in the adaptive-training, validation, and independent test subsets for both the ideal-MHD and ideal-plus-diamagnetic models for the seed-456 realization. Their close agreement indicates that the validation and independent test subsets span a similar range of stability behavior to the adaptive-training pool within this realization.

Figure~\ref{fig:gp_results} compares the Gaussian Process predictions of $\widehat{\gamma}_{\max}$ with the corresponding \textsc{BOUT++} results. Final test metrics were first calculated separately for each of the five random data-partition realizations and were then summarized using their mean and sample standard deviation. For the ideal-MHD model, the surrogate achieved $R^2=0.978\pm0.013$, RMSE $=0.0153\pm0.0046$, MAE $=0.00938\pm0.00059$, and MRE $=(4.90\pm0.24)\%$. For the ideal-plus-diamagnetic model, it achieved $R^2=0.966\pm0.009$, RMSE $=0.00887\pm0.00092$, MAE $=0.00588\pm0.00020$, and MRE $=(9.17\pm0.62)\%$. The displayed parity points correspond to the designated seed-456 realization, whereas the numerical metrics summarize performance across all five realizations.

Figure~\ref{fig:growth_spectrum_example} compares the predicted and simulated growth-rate spectra for both physics models. To select an objective example, the spectral RMSE was calculated for each independent-test equilibrium $i$ and physics model $m$ as
\begin{equation}
\mathrm{RMSE}_{i,m}
=
\left[
\frac{1}{N_n}
\sum_{j=1}^{N_n}
\left(
{\widehat{\gamma}^{\,\mathrm{GP}}_{i,m}(n_j)
-
\widehat{\gamma}_{i,m}(n_j)}
\right)^2
\right]^{1/2},
\label{eq:model_spectral_rmse}
\end{equation}
where $m\in\{\mathrm{ideal},\mathrm{dia}\}$ denotes the ideal-MHD or ideal-plus-diamagnetic model, $N_n=16$, and $n_j=5,10,\ldots,80$. A combined spectral RMSE was then defined by assigning equal weight to the two physics models:
\begin{equation}
\mathrm{RMSE}_{i,\mathrm{combined}}
=
\left[
\frac{
\mathrm{RMSE}_{i,\mathrm{ideal}}^2+
\mathrm{RMSE}_{i,\mathrm{dia}}^2
}{2}
\right]^{1/2}.
\label{eq:combined_spectral_rmse}
\end{equation}
Because both physics models contain 16 output channels, Eq.~\eqref{eq:combined_spectral_rmse} is equivalent to calculating the RMSE over all 32 mode-resolved outputs. The equilibrium whose combined spectral RMSE was closest to the median of the seed-456 independent-test distribution was selected for Fig.~\ref{fig:growth_spectrum_example}. The selected equilibrium
has $\mathrm{RMSE}_{\mathrm{combined}}=7.115\times10^{-3}$, compared with a test-set median of $7.109\times10^{-3}$. Its model-specific spectral errors are $\mathrm{RMSE}_{\mathrm{ideal}}=9.517\times10^{-3}$ and $\mathrm{RMSE}_{\mathrm{dia}}=3.267\times10^{-3}$.

For this median-error equilibrium, the Gaussian Process reproduces the overall spectral shape and the dominant unstable toroidal mode for both physics models. The associated latent posterior standard deviations provide a mode-resolved indication of surrogate uncertainty.


Although Fig.~\ref{fig:growth_spectrum_example} illustrates the prediction quality for one objectively selected equilibrium, it does not show how the surrogate error varies with toroidal mode number over all independent test subsets. To quantify this dependence, the mode-resolved test RMSE was calculated separately for each physics model and each of the five random data-partition realizations. For realization $s$, physics model $m$, and toroidal mode number $n_j$, the RMSE was defined as
\begin{equation}
\mathrm{RMSE}_{m}^{(s)}(n_j)
=
\left[
\frac{1}{N_{\mathrm{test}}^{(s)}}
\sum_{i=1}^{N_{\mathrm{test}}^{(s)}}
\left(
{\widehat{\gamma}^{\,\mathrm{GP},(s)}_{i,m}(n_j)
-
\widehat{\gamma}^{(s)}_{i,m}(n_j)}
\right)^2
\right]^{1/2},
\label{eq:mode_resolved_rmse}
\end{equation}
where $m\in\{\mathrm{ideal},\mathrm{dia}\}$,
$n_j=5,10,\ldots,80$, and $N_{\mathrm{test}}^{(s)}$ is the number of equilibria in the independent test subset for realization $s$.

The values of $\mathrm{RMSE}_{m}^{(s)}(n_j)$ were first calculated separately for each realization and were then summarized by their mean and sample standard deviation over the five realizations:
\begin{align}
\overline{\mathrm{RMSE}}_{m}(n_j)
&=
\frac{1}{5}
\sum_{s=1}^{5}
\mathrm{RMSE}_{m}^{(s)}(n_j),
\label{eq:mode_rmse_seed_mean}
\\
\sigma_{\mathrm{RMSE},m}(n_j)
&=
\left[
\frac{1}{4}
\sum_{s=1}^{5}
\left(
\mathrm{RMSE}_{m}^{(s)}(n_j)
-
\overline{\mathrm{RMSE}}_{m}(n_j)
\right)^2
\right]^{1/2}.
\label{eq:mode_rmse_seed_std}
\end{align}

\begin{figure*}[t]
    \centering
    \includegraphics[width=0.68\textwidth]{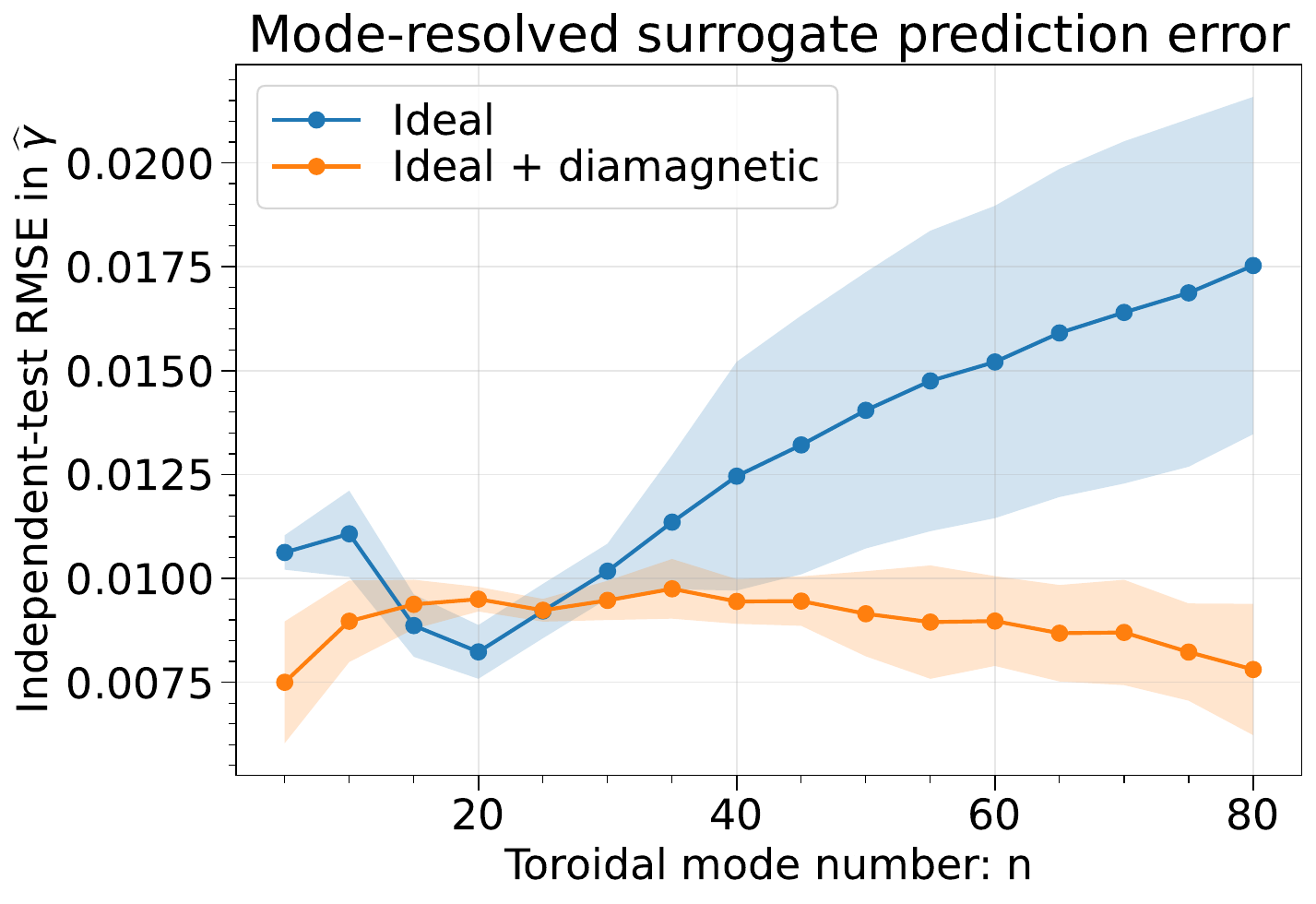}
    \caption{\textbf{Mode-resolved Gaussian Process prediction error on the independent test subsets.}
    The independent-test RMSE of the normalized growth rate $\widehat{\gamma}$ is shown as a function of toroidal mode number for the ideal-MHD and ideal-plus-diamagnetic models. For each mode number and physics model, the RMSE was first calculated over the independent test equilibria within each realization. The solid curves show the mean of these RMSE values over the five random data-partition realizations, and the shaded regions denote one sample standard deviation across the realizations.}
    \label{fig:mode_resolved_rmse}
\end{figure*}

Figure~\ref{fig:mode_resolved_rmse} shows the resulting mode-resolved test errors. For the ideal-MHD model, the mean RMSE generally increases toward the higher toroidal mode numbers, with a corresponding increase in variability across the five realizations. In contrast, the ideal-plus-diamagnetic model exhibits a comparatively weak dependence on mode number, with its largest errors occurring primarily at intermediate $n$. These results complement the aggregate test metrics by identifying the spectral regions in which the surrogate predictions are most sensitive to the particular data partition and training realization.

Taken together, these results demonstrate that the Gaussian Process surrogate accurately reproduces the scalar stability metrics and the principal features of the two sixteen-mode growth-rate spectra generated by \textsc{BOUT++}, establishing the surrogate as a rapid approximation to the corresponding \textsc{BOUT++} linear-stability calculations within the parameter domain represented by the training database.

\subsection{Prediction Errors, Uncertainty Calibration, and Computational Performance}
\label{sec:errors}


\begin{figure*}[t]
    \centering
    \includegraphics[width=0.95\textwidth]{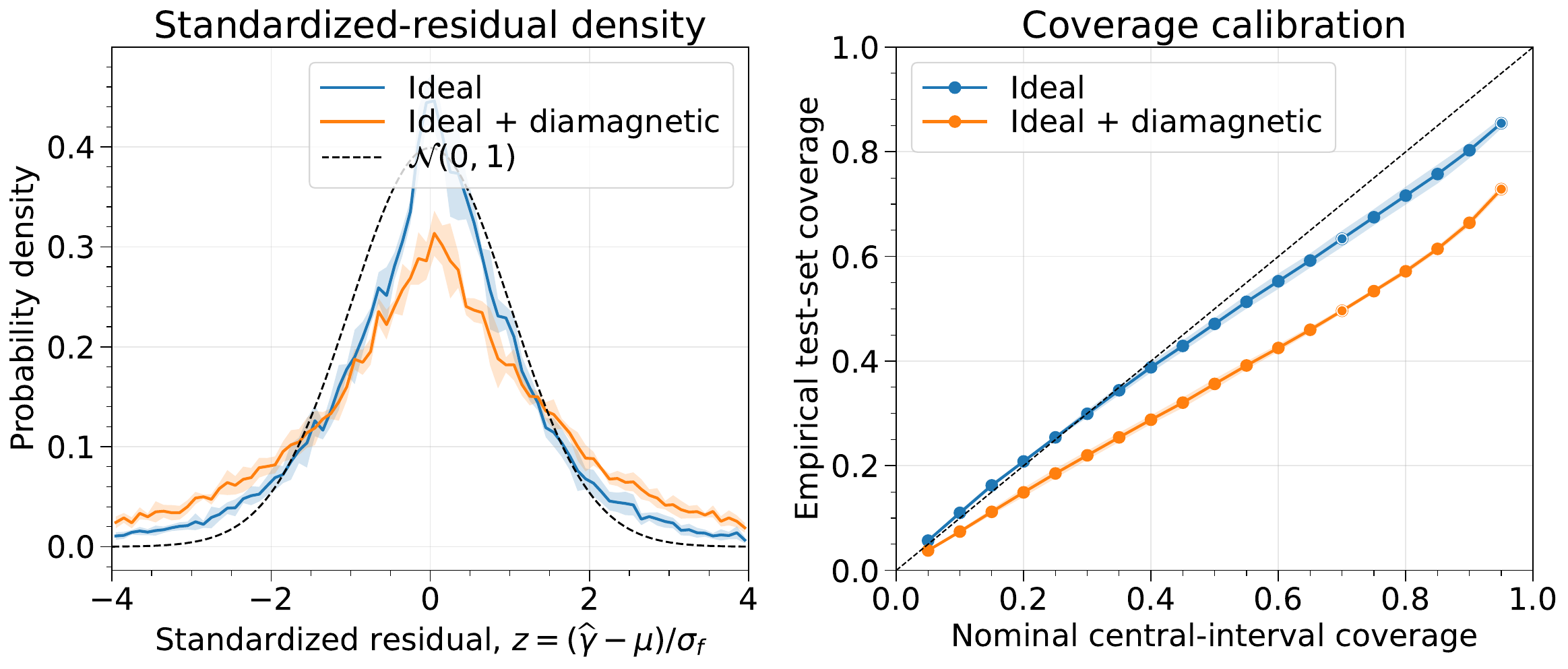}
    \caption{\textbf{Calibration of the Gaussian Process latent posterior uncertainty on the independent test subsets.}
    \textbf{(a)} Probability-density distributions of the standardized residual
    $z=(\widehat{\gamma}-\mu)/\sigma_f$
    for the ideal-MHD and ideal-plus-diamagnetic models. Within each realization, the residuals are pooled over all independent-test equilibria and all sixteen toroidal-mode outputs for the corresponding physics model. The dashed curve denotes the standard-normal density $\mathcal{N}(0,1)$. Solid curves show the mean density over the five random data-partition realizations, and shaded regions denote one sample standard deviation.
    \textbf{(b)} Empirical test-set coverage as a function of nominal Gaussian central-interval coverage. The diagonal dashed line represents perfect calibration. The curves show the mean empirical coverage over the five realizations, and the shaded regions indicate one sample standard deviation. Curves below the diagonal indicate that the latent posterior intervals are too narrow.}
    \label{fig:posterior_calibration}
\end{figure*}

The uncertainty estimates returned by the Gaussian Process were evaluated on the independent test subsets of the five random data-partition realizations. The quantity used here is the latent-function posterior standard deviation, $\sigma_f$, which excludes the additional Gaussian-likelihood observation-noise contribution. For each physics model, the test predictions were pooled over all independent-test equilibria and all sixteen toroidal-mode outputs within each realization. The standardized residual for prediction $i$, physics model $m$, and realization $s$ was calculated as
\begin{equation}
z_{i,m}^{(s)}
=
\frac{
\widehat{\gamma}^{(s)}_{i,m}
-
\mu_{i,m}^{(s)}
}{
\sigma_{f,i,m}^{(s)}
},
\label{eq:standardized_residual}
\end{equation}
where \(\mu_{i,m}^{(s)}\) and \(\sigma_{f,i,m}^{(s)}\) are the Gaussian Process latent posterior mean and standard deviation, respectively. For well-calibrated Gaussian posterior uncertainties, the standardized residuals should approximately follow a standard
normal distribution. Calibration was additionally assessed using the empirical coverage of central posterior intervals. For a nominal central coverage \(p\), the corresponding Gaussian multiplier is
\begin{equation}
q_p
=
\Phi^{-1}\left(\frac{1+p}{2}\right),
\label{eq:central_interval_multiplier}
\end{equation}
where \(\Phi^{-1}\) is the inverse standard-normal cumulative
distribution function. The empirical coverage for realization \(s\)
and physics model \(m\) was calculated as
\begin{equation}
C_m^{(s)}(p)
=
\frac{1}{N_m^{(s)}}
\sum_{i=1}^{N_m^{(s)}}
\mathbb{I}
\left[
\left|
\widehat{\gamma}^{(s)}_{i,m}
-
\mu_{i,m}^{(s)}
\right|
\leq
q_p \sigma_{f,i,m}^{(s)}
\right],
\label{eq:empirical_coverage}
\end{equation}
where \(\mathbb{I}[\cdot]\) is the indicator function and
\(N_m^{(s)}\) is the total number of test predictions for model \(m\)
in realization \(s\). The coverage was evaluated separately for each
realization, after which its mean and sample standard deviation were
calculated over the five realizations.

As shown in Fig.~\ref{fig:posterior_calibration}, the standardized-residual distributions are broader than the standard-normal reference, and the empirical-coverage curves generally lie below the ideal-calibration line. At the nominal $68.3\%$ central interval, the empirical coverage is $61.97\pm1.54\%$ for the ideal-MHD model and $48.34\pm0.50\%$ for the ideal-plus-diamagnetic model. At the nominal $95.4\%$ interval, the corresponding empirical coverages are $86.01\pm0.90\%$ and $73.56\pm0.67\%$, respectively. The latent posterior uncertainties are therefore underdispersed, particularly for the ideal-plus-diamagnetic model. Consequently, the posterior standard deviations provide useful relative surrogate-uncertainty indicators and acquisition scores, but they should not be interpreted as fully calibrated predictive intervals.

\begin{figure*}[t]
  \centering

  \begin{overpic}[width=0.9\textwidth]{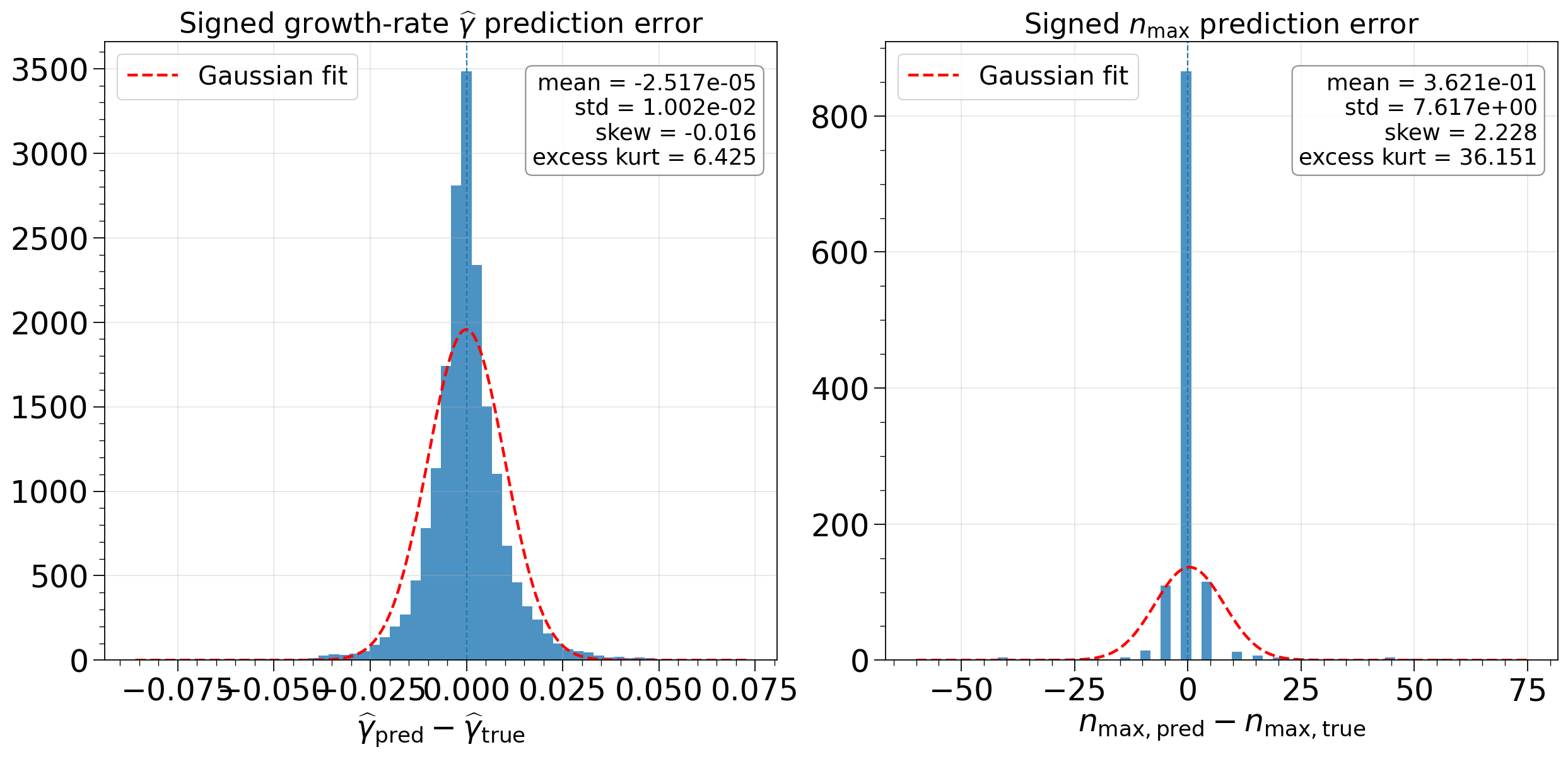}
    \put(8,8){\large\textbf{(a)}}
    \put(58,8){\large\textbf{(b)}}
  \end{overpic}
  
  \caption{\textbf{Prediction-error distributions of the Gaussian Process surrogate on the independent test set.}
  \textbf{(a)} Distribution of the signed prediction errors for all mode-resolved linear growth-rate predictions, $\widehat\gamma_{\rm pred}-\widehat\gamma_{\rm true}$, aggregated over the independent test equilibria for both the ideal-MHD and ideal-plus-diamagnetic models (32 surrogate output quantities per equilibrium). The red dashed curve shows a Gaussian fit to the error distribution. The errors are centered near zero, with a mean of $-2.517\times10^{-5}$ and negligible skewness (-0.016), indicating that the surrogate is essentially unbiased. The distribution exhibits an approximately Gaussian core with moderately heavier tails, corresponding to an excess kurtosis of 6.425.
  \textbf{(b)} Distribution of the signed prediction error for the dominant unstable toroidal mode, $n_{\max,\mathrm{pred}}-n_{\max,\mathrm{true}}$.  The pronounced peak at zero demonstrates that the surrogate correctly predicts the dominant unstable toroidal mode for the majority of independent test cases. The overall signed-error distribution exhibits positive skewness (2.23), arising from a small number of relatively large positive deviations. The exact-match rates are 85.17\% and 64.14\% for the ideal-MHD and ideal-plus-diamagnetic models, respectively, while 95.69\% and 96.90\% of the corresponding predictions fall within two neighboring modes.
  }
  \label{fig:error_distribution}
\end{figure*}

The prediction-error statistics in Fig.~\ref{fig:error_distribution} provide a complementary assessment of surrogate performance on the designated seed-456 independent test subset. Figure~\ref{fig:error_distribution}(a) summarizes the signed errors for all mode-resolved linear growth-rate predictions. The histogram pools the prediction errors from the complete growth-rate spectra for both the ideal-MHD and ideal-plus-diamagnetic models, corresponding to 32 surrogate output quantities for each independent test equilibrium. The errors are centered close to zero, with a mean of approximately $-2.5\times10^{-5}$ and skewness of $-0.016$. The distribution exhibits an approximately Gaussian core with heavier tails, corresponding to an excess kurtosis of 6.425. Thus, the pooled error distribution exhibits negligible overall bias, although its heavy tails indicate the presence of a comparatively small number of larger prediction errors.

Figure~\ref{fig:error_distribution}(b) summarizes the signed prediction errors for the dominant unstable toroidal mode number, $n_{\max}$. A pronounced peak at zero indicates that the surrogate correctly identifies the dominant mode for the majority of the seed-456 independent test cases. The pooled signed-error distribution has a positive skewness of 2.23, reflecting a small number of comparatively large positive deviations. For the ideal-MHD model, the exact-match rate is 85.17\%, while 93.97\% and 95.69\% of the predictions fall within one and two neighboring sampled modes, respectively. For the \textsc{BOUT++} mode spacing $\Delta n=5$, these conditions correspond to $|\Delta n_{\max}|\leq5$ and $|\Delta n_{\max}|\leq10$, respectively. The mean absolute error in $n_{\max}$ is 2.26. For the ideal-plus-diamagnetic model, the exact-match rate is 64.14\%, while 94.14\% and 96.90\% of the predictions fall within one and two neighboring sampled modes, respectively. The corresponding mean absolute error is 2.59. Although the ideal-plus-diamagnetic exact-match rate is lower, more than 93\% of the predictions for both physics models remain within one neighboring sampled mode of the \textsc{BOUT++} value.

A complete \textsc{BOUT++} linear-stability scan for a single equilibrium consists of 32 separate calculations: sixteen toroidal mode numbers for each of the ideal-MHD and ideal-plus-diamagnetic physics models. The complete scan requires approximately 21~min using 128 CPU cores, corresponding to approximately 44.8 core-hours per equilibrium. In contrast, the trained Gaussian Process surrogate predicts all 32 output quantities in approximately 20~ms on a single CPU core, corresponding to $5.56\times10^{-6}$ core-hours. The surrogate therefore provides an approximately $6.3\times10^{4}$-fold wall-clock speedup and an approximately $8.1\times10^{6}$-fold reduction in computational cost. These comparisons exclude the one-time costs associated with simulation database generation and surrogate training.

This reduction in prediction time and computational cost enables rapid evaluation of pedestal linear stability within the parameter domain represented by the present simulation database.

\section{Discussion and Limitations}
\label{sec:discussion}

\begin{figure*}[t]
  \centering
  \includegraphics[width=0.68\textwidth]{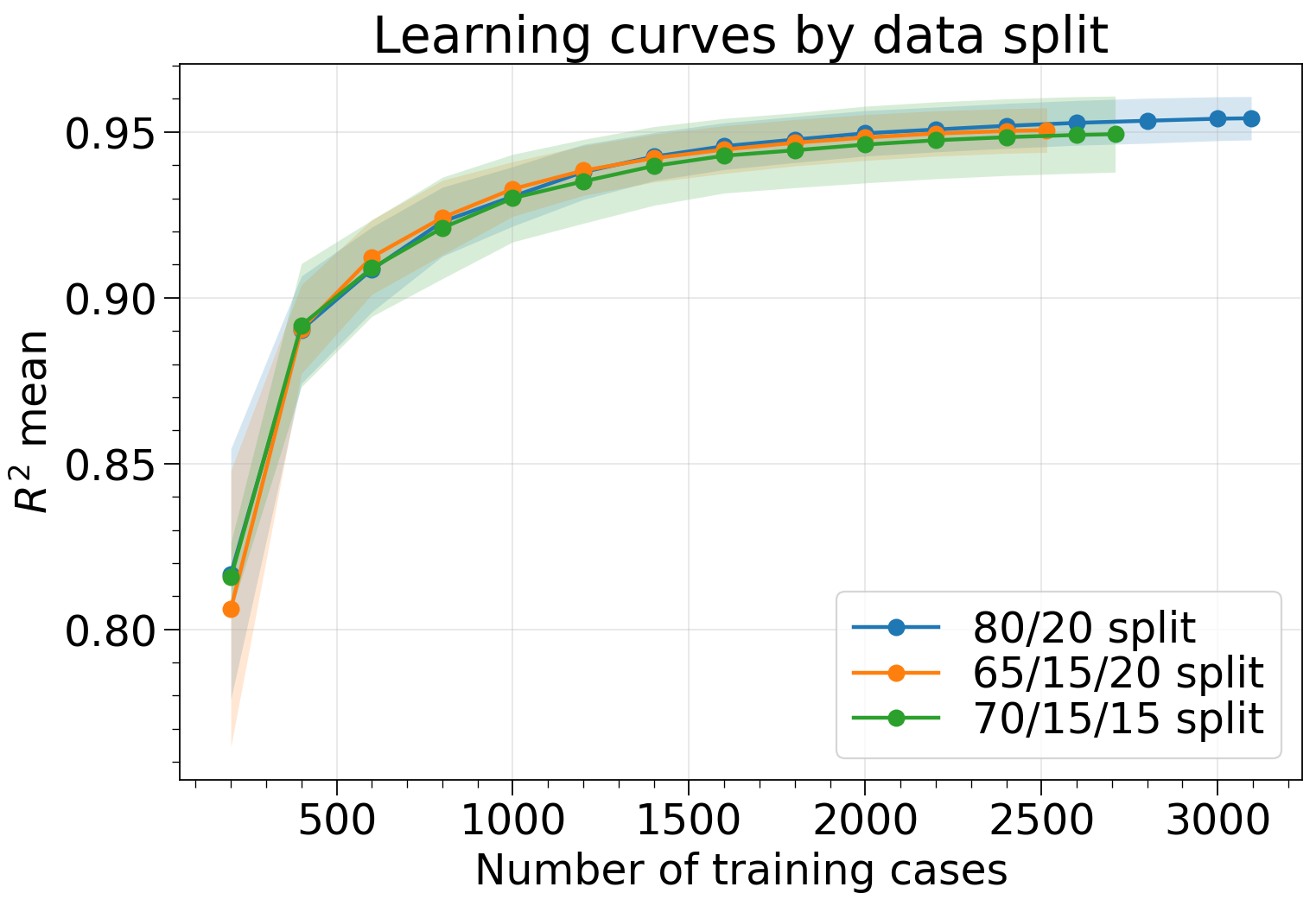}
  \caption{\textbf{Sensitivity of surrogate learning performance to the data-partition strategy.}
  Mean validation $R^2$ is shown as a function of training-set size for the 80/20 training/validation partition, the 65/15/20 training/validation/test partition, and the 70/15/15 training/validation/test partition adopted for the main analysis. Shaded regions denote one sample standard deviation over five random data-partition realizations. The three learning curves exhibit similar convergence behavior and validation accuracy over their common training-size range, indicating that the principal conclusions regarding surrogate convergence are robust to the choice of data partition.}
  \label{fig:data_split}
\end{figure*}

The present results demonstrate that Gaussian Process Regression provides an accurate and computationally efficient approximation to \textsc{BOUT++} linear pedestal-stability calculations within the parameter domain represented by the present single-shape DIII-D database. The surrogate reproduces the principal features of the two sixteen-mode growth-rate spectra corresponding to the ideal-MHD and ideal-plus-diamagnetic physics models, yielding 32 surrogate output quantities for each equilibrium. It also accurately predicts the maximum linear growth rate and generally identifies the dominant unstable toroidal mode.

The prediction accuracy is nevertheless target dependent. The present surrogate performs well for the maximum linear growth rate and dominant toroidal mode over most of the represented domain, whereas larger errors occur in the weak-growth-rate regime. In addition, Fig.~\ref{fig:mode_resolved_rmse} shows that the ideal-MHD prediction error increases toward the highest toroidal mode numbers. This trend is consistent with the reduced database coverage at the largest values of the normalized pressure-gradient and current-density parameters shown in Fig.~\ref{fig:growth_database}, where high-$n$ ideal-MHD behavior is more strongly represented. The observed increase in high-$n$ error may therefore reflect, at least in part, the smaller number of training equilibria available in this region of parameter space. Because accurate resolution of small positive growth rates and high-$n$ behavior is important for pedestal-stability assessment, the reported global performance metrics should not be interpreted as demonstrating uniform accuracy across all growth-rate regimes or throughout the realized parameter domain. Boundary-focused validation and targeted sampling will be needed to improve prediction accuracy in these regions.

The learning curves show rapid improvement as the training set is expanded, followed by diminishing gains beyond approximately 2,000 training equilibria. Within the existing simulation database, uncertainty-guided sample selection provides a modest improvement in sample efficiency relative to random selection over intermediate training-set sizes. The two strategies exhibit essentially the same performance at the initial training size because they begin from the same randomly selected subset, and their performance converges as the training-set size approaches the full adaptive-training pool. The benefit of uncertainty-guided selection is therefore primarily associated with reaching a given level of validation accuracy using fewer training samples rather than improving the final converged accuracy.

The overlap between the variability bands indicates that this improvement is modest relative to realization-to-realization variability. This behavior is consistent with the realized database, in which the populated regions remain sufficiently redundant that uncertainty-guided selection offers only limited advantage within the existing candidate pool. It does not, however, imply that the remaining prediction error arises solely from database coverage. Additional contributions may originate from the selected input representation, the Gaussian Process covariance kernel, the independent-output formulation, and unresolved correlations among neighboring toroidal modes and between the two physics models.

The Gaussian Process latent posterior uncertainty provides a relative indicator of regions that are less strongly constrained by the existing training data. In the present work, this information is used only for pool-based adaptive sample selection among previously computed equilibria. No new \textsc{Varyped} equilibria, \textsc{Hypnotoad} meshes, or \textsc{BOUT++} calculations are generated automatically. The adaptive results should therefore be interpreted as a demonstration of uncertainty-guided training-set construction rather than a completed closed-loop adaptive simulation capability.

The present acquisition score is based on the mean Gaussian Process latent posterior standard deviation across the 32 output channels and is intended to identify comparatively uncertain cases within the existing candidate pool. It is not specifically formulated to resolve the stable--unstable boundary. Future work should compare this criterion with information-gain, boundary-focused, and other physics-informed acquisition objectives, depending on whether the goal is improved global spectral accuracy or more precise identification of marginal stability.

The following considerations define the domain of validity and appropriate interpretation of the present surrogate.

\textit{Geometric coverage.}
The database retains a single, fixed DIII-D boundary shape. Although the scanned pressure, current-density, and safety-factor parameters span a broad range within this configuration, the surrogate has not been tested across variations in elongation, triangularity, squareness, magnetic topology, or machine size. Its predictions should therefore not be extrapolated to substantially different plasma shapes or devices without additional training data and validation.

\textit{Realized parameter-space coverage.}
Although the requested \textsc{Varyped} scan was constructed from discrete, approximately uniform sequences of input parameters, the resulting distribution is nonuniform in the equilibrium-derived features used for surrogate training. Equilibrium-convergence limitations and quality-control exclusions further reduce coverage in some regions. The surrogate is therefore validated only within the successfully realized simulation domain, and its posterior uncertainty should not be interpreted as guaranteeing reliable extrapolation into sparsely sampled or excluded regions.

\textit{Physics scope.}
The target data are obtained from linear \textsc{BOUT++} calculations using the ideal-MHD and ideal-plus-diamagnetic physics models. The surrogate predicts the corresponding linear growth rates rather than nonlinear ELM amplitudes, energy losses, transport dynamics, or experimentally observed ELM onset. Finite resistivity, additional drift physics, equilibrium rotation, nonlinear mode coupling, and plasma--wall interactions remain outside the demonstrated scope.

\textit{Surrogate representation.}
The surrogate consists of independent single-task Gaussian Processes for the 32 surrogate output quantities. This formulation is straightforward and provides a latent posterior uncertainty estimate for each output, but it does not explicitly learn correlations among neighboring toroidal modes or between the two physics models. Multi-output Gaussian Processes, reduced spectral representations, neural operators, and other structured surrogate approaches may become advantageous as the database expands to include mode structures, nonlinear fields, and higher-dimensional outputs.

\textit{Uncertainty interpretation.}
The latent posterior uncertainties are conditional on the selected input features, covariance kernel, likelihood model, training distribution, and independent-output formulation. They characterize uncertainty within the surrogate model and should not be interpreted as a complete measure of physical, numerical, or experimental uncertainty. The calibration diagnostics in Fig.~\ref{fig:posterior_calibration} show that the posterior uncertainties are underdispersed, particularly for the ideal-plus-diamagnetic model. The posterior standard deviations therefore provide useful relative indicators for pool-based sample acquisition, but they should not be interpreted as fully calibrated predictive intervals. Moreover, the present calibration analysis is restricted to in-domain independent test subsets and does not establish reliable uncertainty behavior under distribution shift. Posterior recalibration, out-of-distribution diagnostics, and comparisons with alternative acquisition criteria should therefore accompany future closed-loop applications.

\textit{Model-development protocol.}
The quality-controlled simulation database is partitioned at the equilibrium level into a 70\% adaptive-training pool, a 15\% validation set, and a 15\% independent test set. The validation set is used during model development for feature selection, learning-curve construction, and adaptive stopping, whereas the independent test set remains excluded from Gaussian Process fitting, adaptive sample acquisition, feature-selection decisions, and stopping. After the feature set and adaptive-training procedure have been fixed, only the performance of the frozen final surrogate is reported as independent-test performance. Repeating this procedure over five random data-partition realizations yields closely consistent learning curves and test performance, indicating that the principal surrogate-accuracy conclusions are not strongly dependent on a particular random partition. This separation between validation and final testing reduces optimistic bias associated with reusing the same held-out samples for both model development and final performance assessment.

The comparable performance obtained on the independent test subsets indicates that the high prediction accuracy observed during validation is retained when the surrogate is evaluated on equilibria that played no role in model-development decisions.

To assess sensitivity to the choice of data partition, the analysis was additionally repeated using an 80/20 training/validation partition and a 65/15/20 training/validation/test partition. Figure~\ref{fig:data_split} compares these results with the 70/15/15 protocol adopted for the main analysis. The three learning curves exhibit closely similar convergence behavior and validation accuracy over their common training-size range, indicating that the principal convergence conclusions are insensitive to reasonable changes in the relative sizes of the training, validation, and test subsets.

\textit{Experimental validation.}
The present work has not yet established experimental validation of the predicted growth-rate spectra or associated stability metrics. Future development should include multiple plasma shapes, additional linear and nonlinear physics models, cross-code comparisons, and validation using experimentally reconstructed equilibria together with observed ELM behavior. These extensions will provide a more demanding assessment of surrogate generalization and clarify whether uncertainty-guided sample selection offers a significant advantage over alternative database-expansion strategies.

\section{Conclusions}
\label{sec:conclusion}

This work presents the focused implementation of ELMO---the Edge Learning and Modeling Orchestrator---as an uncertainty-aware simulation-to-surrogate workflow for rapid pedestal linear-stability prediction. The workflow integrates systematic equilibrium generation using \textsc{Varyped}/BOUT\_DB, field-aligned mesh generation using \textsc{Hypnotoad}, large-scale \textsc{BOUT++} linear-stability calculations, automated campaign execution and reduced-data extraction, physics-informed feature selection, and Gaussian Process Regression.

Of the 7,992 requested equilibrium configurations for a single DIII-D plasma shape, 3,901 converged during EFIT reconstruction, 3,877 produced valid \textsc{Hypnotoad} meshes, and 3,869 completed the subsequent \textsc{BOUT++} linear-stability calculations without identified numerical issues and satisfied the quality-control criteria for surrogate-model development. Each retained equilibrium contributes two sixteen-mode growth-rate spectra corresponding to the ideal-MHD and ideal-plus-diamagnetic physics models, yielding 123,808 high-fidelity \textsc{BOUT++} linear-stability calculations and 32 surrogate output quantities per equilibrium.

Using eight equilibrium-derived pedestal features, the Gaussian Process surrogate accurately reproduces the principal features of the two sixteen-mode growth-rate spectra together with the maximum linear growth rate and dominant unstable toroidal mode. Model development is performed using separate adaptive-training and validation subsets, while an independent 15\% test subset within each of five random data-partition realizations is reserved for final performance assessment. For the maximum linear growth rate, the five-realization test results are $R^2=0.978\pm0.013$ for the ideal-MHD model and $R^2=0.966\pm0.009$ for the ideal-plus-diamagnetic model, where the reported values denote the mean and sample standard deviation. The mode-resolved test analysis further shows that the ideal-MHD error increases toward the highest toroidal mode numbers, identifying a region in which additional targeted sampling may improve surrogate performance.

A complete prediction of all 32 surrogate output quantities requires approximately 20~ms on a single CPU core, compared with approximately 21~min using 128 CPU cores for the corresponding \textsc{BOUT++} scan. The surrogate therefore provides an approximately $6.3\times10^{4}$-fold wall-clock speedup and an approximately $8.1\times10^{6}$-fold reduction in computational cost. These comparisons exclude the one-time costs associated with simulation database generation and surrogate training.

Pool-based uncertainty-guided sample selection within the existing simulation database provides a modest improvement in sample efficiency relative to random selection over intermediate training-set sizes, while the two strategies converge as the available training pool is exhausted. Calibration diagnostics show that the latent posterior uncertainties are underdispersed, particularly for the ideal-plus-diamagnetic model. They therefore provide useful relative acquisition scores but should not be interpreted as fully calibrated predictive intervals. The adaptive component demonstrated here is restricted to sample selection among previously computed equilibria; no new \textsc{Varyped} equilibria, \textsc{Hypnotoad} meshes, or \textsc{BOUT++} simulations are generated automatically. The present implementation therefore establishes the simulation, data-reduction, surrogate-modeling, and uncertainty-aware sample-selection components required for future adaptive database development. Extensions to multiple plasma shapes, additional linear and nonlinear physics models, fully closed-loop adaptive simulation workflows, and experimental validation remain subjects for future work.


\section*{Acknowledgements}

This work was performed under the auspices of the U.S. Department of Energy by Lawrence Livermore National Laboratory under Contract DE-AC52-07NA27344, LLNL-JRNL-2022077 and was supported by the SciDAC ABOUND Project, SCW1832.
This research used resources of the National Energy Research Scientific Computing Center (NERSC), a DOE Office of Science User Facility supported by the Office of Science of the U.S. Department of Energy
under Contract No. DE-AC02-05CH11231 using NERSC award FES-ERCAP0026742.


\section*{Data Availability}

The data that support the findings of this study are available from the
corresponding author upon reasonable request.


\appendix
\section{Software Environment and Reproducibility}
\label{app:software}

\begin{table*}
\caption{Software environment used for database generation, workflow execution,
and surrogate analysis. Version numbers, commit identifiers, or availability status describe the software environment used to produce the reported results.
}
\label{tab:software_environment}
\centering

\renewcommand{\arraystretch}{1.12}
\setlength{\tabcolsep}{5pt}

\begin{ruledtabular}
\begin{tabular}{lll}
\textbf{Software component} &
\textbf{Version or availability} &
\textbf{Role in the ELMO workflow} \\
\hline

\textsc{Varyped}/\texttt{BOUT\_DB}
& \texttt{\makecell[l]{Internal GA\\software}}
& \parbox[t]{0.53\textwidth}{Generation of the structured pedestal-equilibrium
database and associated equilibrium-derived quantities.} \\

EFIT
& \texttt{\makecell[l]{Internal GA\\software}}
& \parbox[t]{0.53\textwidth}{Reconstruction of self-consistent
magnetohydrodynamic equilibria for the requested \textsc{Varyped}
parameter combinations.} \\

\textsc{Hypnotoad}
& \texttt{[idl,d7c62a1]}
& \parbox[t]{0.53\textwidth}{Generation of field-aligned computational
meshes from the converged equilibria using the automated headless grid-generation workflow.} \\

IDL
& \texttt{[idl/8.9]}
& \parbox[t]{0.53\textwidth}{Execution environment for the automated IDL Hypnotoad
grid-generation workflow.} \\

\textsc{BOUT++}
& \texttt{[v5.1.1-813-g385955543]}
& \parbox[t]{0.53\textwidth}{Mode-resolved linear-stability calculations
using the ideal-MHD and ideal-plus-diamagnetic physics models.} \\

EFFIS
& \texttt{[2.0.0]}
& \parbox[t]{0.53\textwidth}{Composition and execution of the simulation
workflow, including run-directory preparation, job submission, monitoring,
and restart of incomplete calculations.} \\

ADIOS
& \texttt{[adios2-2.10.1]}
& \parbox[t]{0.53\textwidth}{Storage and transport of raw simulation
outputs and reduced analysis products.} \\

HPC Campaign
& \texttt{[0.7.1]}
& \parbox[t]{0.53\textwidth}{Management of campaign metadata and
associations among equilibria, meshes, simulation configurations,
execution status, and reduced outputs.} \\

Python
& \texttt{[python/3.13]}
& \parbox[t]{0.53\textwidth}{Post-processing, feature extraction,
quality control, surrogate training, and performance evaluation.} \\

PyTorch
& \texttt{[2.11.0]}
& \parbox[t]{0.53\textwidth}{Tensor operations and numerical backend
for Gaussian Process training and inference.} \\

GPyTorch
& \texttt{[1.15.2]}
& \parbox[t]{0.53\textwidth}{Implementation and optimization of the
independent single-task Gaussian Process models.} \\

BoTorch
& \texttt{[0.17.2]}
& \parbox[t]{0.53\textwidth}{Probabilistic-model utilities and
uncertainty-guided acquisition within the adaptive-training pool.} \\

\end{tabular}
\end{ruledtabular}
\end{table*}

The reduced dataset supporting the findings of this study comprises the equilibrium-derived surrogate inputs, quality-control metadata, data-partition indices, and mode-resolved BOUT++ linear growth rates used for surrogate training and evaluation. Subject to applicable institutional review and release requirements, these data may be obtained from the corresponding author upon reasonable request.

The reduced-data processing and surrogate-analysis stages can be reproduced using the derived dataset, workflow configurations, data-reduction and surrogate-analysis scripts, and documented software environment. Regeneration of the underlying equilibrium database, however, requires access to Varyped/BOUT\_DB and the General Atomics implementation of EFIT, which are internal research software packages and are not publicly distributed. The publicly available software dependencies, including BOUT++, Hypnotoad, GPyTorch, and BoTorch, together with the versions or revisions used in this study, are summarized in Table~\ref{tab:software_environment}. The ELMO workflow configurations, data-reduction scripts, and surrogate-analysis scripts may also be obtained from the corresponding author upon reasonable request, subject to applicable institutional review and software-release requirements.


\section*{References}
\bibliographystyle{apsrev4-2}
\bibliography{references}


\end{document}